\documentclass[10pt]{article}
\usepackage{graphicx}
\date{November 1999}

\title{ {\LARGE\bf Muon's Behaviors under
Bremsstrahlung  with both the LPM effect and the
Ter-Mikaelian effect and Direct Pair Production with the LPM effect}}
\author{ S.~Polityko$^1$, N.~Takahashi$^2$,
M.~Kato$^3$, Y.~Yamada$^4$ and A.~Misaki$^5$ \\
{\it $^1$Irkutsk State University, Irkutsk, 664003 Russia }\\
{\it $^2$Facutly of Science and Technology, Hirosaki University,}\\
{\it Hirosaki 036-8561 Japan }\\
{\it $^3$Kyowa Interface Science Co.Ltd., Asaka, 351-0033 Japan }\\
{\it $^4$Hitachi Goverment and Public Cooperation System}\\
{\it Engineering Ltd., Koto-ku, Tokyo, 136-8633 Japan }\\
{\it $^5$National Graduate Institute for Policy Studies,}\\
{\it 2-2 Wakamatsu-cho, Shinjyuko-ku, 162-8677}\\
{\it and Institute for Policy Science, Saitama University,}\\
{\it Urawa 338-8570 Japan }}

\begin{document}
\maketitle

\begin{abstract}
Differential and integral cross sections of the muon are calculated
in the materials: water, standard rock, iron and lead with and without
the LPM effect. The corresponding cross sections are also calculated with
dielectric supression effect (Ter-Mikaelian effect), in addition to the
LPM effect.
In our calculations the LPM effect for muon is provided to be effective
from $10^{14}$ eV to $10^{20}$ eV depending on materials in the
bremsstrahlung process, while it is provided to be completely negligible
in the direct pair production process up to $10^{22}$ eV even for lead.

        As for the dielectric suppression effect it is only effective
between $10^{12}$ eV and $10^{20}$ eV in the case with the LPM effect,
while it is effective above $10^{12}$ eV in the case without LPM
effect.
        To demonstrate the importance of the LPM effect in the bremsstrahlung
process, depth intensity relation of muon, energy spectrum, range distribution
 and survival
probability are calculated taking into account bremsstrahlung spectrum.

\end{abstract}
%%%%%%%%%%%%%%%%%%%%%%%%%%%%%%%%%%%%%%%%%%%%%%%%
\vspace{0.5cm}
\section{Introduction}
\vspace{0.5cm}
%%%%%%%%%%%%%%%%%%%%%%%%%%%%%%%%%%%%%%%%%%%%%%%%

     In the early 1950's , Landau , Pomeranchuk and Migdal realized that
because of the low longitudinal momentum transfer between the nucleus and
the fast particle, bremsstrahlung is not instantaneous, but occurs over a
finite formation zone. During this time, external influences can perturb
the fast particle and suppress the photon emission. When this happens, the
usual Bethe-Heitler formula fails. Initially, Landau and Pomeranchuk studied
suppression by multiple scattering with semiclassical argument \cite{LD}.
Later, Migdal presented a fully quantum treatment\cite{Mig}. Now this
phenomenon is well known as the LPM effect.
Cosmic ray physicists have been interested in the LPM  effect  which becomes
effective  at extremely high energies. Besides the LPM effect is expected
to change drastically the structure of the electromagnetic cascade showers
at extremely high energies, the structures of which are closely related to
their physical interpretation. Up to now, the experimental proofs of the LPM
effect have been tried by many investigators in both cosmic rays
\cite{Fow} and accelerators \cite{Var},\cite{Perl}.

Recently, the existence of the LPM effect has been finally confirmed by the SLAC
electron accelerator \cite{Ant1} and the longitudinal effect \cite{Ant2}
has been also reconfirmed \cite{Ant3}.
     Electromagnetic cascade shower with the LPM effect is called the LPM
shower, while the LPM effect, namely that at not-extremely high energies is
called the BH (Bethe-Heitler) shower.  The LPM shower has two distinguished
characteristics over the BH shower: a) The average behaviors of the LPM
shower are quite different from those of the BH shower \cite{Mis} and b) The
individual behaviors of the LPM shower are quite average behaviors of
the LPM showers \cite{Kon}.

     From the first characteristics of the LPM shower, we could easily
understand that the size of the experimental apparatus for extremely high
energy phenomena required should be much larger than that in usually imagined,
while from the second characteristics of the LPM shower we understand that
we should create a new method to analyze  extremely high energy phenomena
which could not be analyzed by the average theory.

     Up to now, we have discussed the characteristics of the LPM effect on
the electrons and photons. On the other hand, the LPM effect due to muon is
also important for extremely high energy phenomena., particularly, for
studies on extremely high energy neutrino astrophysics which are quite
an uncultivated field. Now, there are two test pilot experiments for high
energy neutrino astrophysics experiment in Baikal \cite{Baikal} and in
Antarctic \cite{Amanda}. In future, we expect to construct the gigantic
extremely
high energy neutrino astrophysics detector, where the principle of the construction
for the apparatus should be different from that of extremely high energy
neutrino astrophysics due to the presence of the LPM effect. For the moment,
 it is more plausible that extremely high energy neutrino could be detected
as extremely high energy muon from the neutrino interaction. In this energy
region, the fluctuations are expected to play a decisivly important role for
the energy determination of muon neutrino at extremely high energy ( over
$10^{15} eV$). The fluctuations in this energy region come from (a) the LPM
effect due to the muon and (b) the LPM effect due to electron
(electromagnetic cascade shower).

     In this paper, we would like to discuss the case (a) which has never
been discussed before. Without exact consideration of the LPM effect in both
muon and electron (photon) we could not determine the energy of the muon at
extremely high energies.
%%%%%%%%%%%%%%%%%%%%%%%%%%%%%%%%%%%%%%%%%%%%%%%%
\vspace{0.5cm}
\section{Theory: derivation of cross-section}
\vspace{0.5cm}
%%%%%%%%%%%%%%%%%%%%%%%%%%%%%%%%%%%%%%%%%%%%%%%%
      At first we remind here the semiclassical picture of LPM.
 The radiation of a relativistic particle in matter develops in a
large region of space along its momentum. The characteristic size of this
region can be easily estimated. When the muon is of a sufficiently high momentum,
$q$  the longitudinal
momentum carried by the virtual photon, becomes very small,
\begin{eqnarray}
q=p_{\mu}-p_{\mu}^{\prime}-k=\sqrt {E_{\mu}^2-m_{\mu}^2 }-
\sqrt {E_{\mu}^{\prime 2}-m_{\mu}^2 } -E_{\gamma },
\label{6}
\end{eqnarray}
\par
\noindent
where $p_{\mu },p_{\mu }^{\prime },E_{\mu }$ and $E_{\mu }^{\prime}$ are
the muon momentum and energy before and after the interaction, respectively,
and $E_{\gamma } $ is the photon energy and $\gamma $is Lorentz factor.
 For high energy muons this
simplifies to
\begin{eqnarray}
q \sim {m_{\mu }^2E_\gamma \over 2E_\mu (E_\mu -E_\gamma )}=
{m_\mu \over 2\gamma } {u\over 1-u},
\end{eqnarray}
\par
\noindent
where $u=E_\gamma /E_\mu $ is the fractional energy of the radiated photon.
This momentum transfer can be very small. Then, by the uncertainty
principle, the virtual photon exchanges the distance $l_c$
\begin{eqnarray}
l_c \sim {\hbar \over q}={2E_{\mu }^2 \over m_{\mu }^2 E_{\gamma }}
\end{eqnarray}

	For example, for a 100 TeV muon emitting a 1 GeV photon, $q$ is
$0.56\cdot 10^{-3}eV/c$ and the coherence length is $3.5\cdot 10^{-2}$ cm.

        The coherence length can also be interpreted as the length in which
stripping of a photon from the muon which radiates it occurs.

	The LPM effect appears when one considers that the muon
must be undisturbed while it traverses this distance. One factor that
can disturb the muon and disrupt the bremsstrahlung is multiple
Coulomb scattering. Semiclassically, if the muon multiple scatters by an angle
$\theta _s$, greater than the angle made by the bremsstrahlung photon,
$\theta _{br} \sim 1/\gamma $, then the bremsstrahlung is suppressed.
The average multiple scattering angle in a nuclear medium is
\begin{eqnarray}
<\theta _{s}^2>=\left({E_s\over E}\right)^2{l\over X_0},
\end{eqnarray}
\par
\noindent
where $E_s =\sqrt{4\pi /\alpha }m=21$ MeV is the characteristic energy,
$l$ is the scatter thickness, and $X_0$ is the radiation length. The LPM
effect becomes important when $\theta _s$ is larger than $\theta _{br}$.
This occurs for

$$
\left({E_s\over E}\right)\sqrt {{l\over X_0}} > {1\over \gamma }
$$

If $l_c$
is larger than the size of an atom, it is necessary to take into account
the interaction of the incident muon not only with the nucleus of the atom
but also with the atomic electrons. This interaction is taken into account as
the screening effect in the Bethe-Heitler theory of bremsstrahlung.If
$l_c$ exceeds the average distance between the atoms of medium,then it is
necessary to take into consideration the influence both of the atomic electrons
and of many atoms on the particle radiation process.

The coherence length
increases with decrease of the frequency of the radiated photon, and for
sufficiently small $E_\gamma $ it can reach macroscopic dimensions.
Therefore, the LPM effect modifies the soft part of bremsstrahlung
spectrum.  For the low energy photon, the photon spectrum is proportional to
$E_{\gamma}^{-1/2}$, in contrast to the $1/E_\gamma $ Bethe-Heitler
spectrum.

     	An analogous effect occurs for pair creation by a high energy photon.
In pair creation, the LPM energy threshold is determined by the lepton with
the low energy. Because of this, the pair creation suppression begins at more
higher energies than bremsstrahlung suppression.

               At low bremsstrahlung photon energies dielectric
suppression effect occurs (also known as the longitudinal
density effect). This influence can be taken into account by introducing
 dielectric permeability $\varepsilon (E_\gamma )$. Ter-Mikaelian
showed, that in the case of soft photons the discount of
dielectric permeability also leads to decrease of bremsstrahlung
probability \cite{TM}. Recently these effects have been observed experimentally
 \cite{Ant3}.
These phenomena are important in the development of electromagnetic
shower accompaning the high energy muon. Thus for the modelling of
electromagnetic showers it is necessary to use the probabilities, including
 LPM effect and the longitudinal density effect. Earlier the discount
of LPM effect for bremsstrahlung and direct pair production by muon
has been carried out in \cite{PTKM}.
%%%%%%%%%%%%%%%%%%%%%%%%%%%%%%%%%%%%%%%%%%%%%%%%
\subsection{Bremsstrahlung of muon in a medium}
\vspace{0.5cm}
%%%%%%%%%%%%%%%%%%%%%%%%%%%%%%%%%%%%%%%%%%%%%%%%
        The appropriate method to consider radiative effects in media
is the quasiclassic approach of considering the processes of bremsstrahlung
and direct pair production
\cite{BKF73},\cite{Pit}. Let
$k=(E_\gamma ,\vec{k}_\gamma)$- be the 4-momentum of bremsstrahlung photons.

        The probability of radiation of a photon by the high energy muon
per a time unit averaged in initial muon polarization and summed in final
particles polarization has the form
\begin{eqnarray}
dW={\alpha \over (2\pi )^2}{d^3k_\gamma \over E_\gamma }{1\over 2T} Re \int_{-T}^{T}
dt\int_0^\infty d\tau \exp(-i\tilde k (x(t+\tau )-x(t))
{\cal L}(\vec{p}(t+\tau),\vec{p}(t)),
\end{eqnarray}

where  ${\cal L}(\vec{p} (t+\tau),\vec{p} (t))$ - is a square of matrix
elements of muon photon radiation,
\begin{eqnarray}
\tilde k_0=E_\mu E_\gamma /(E_\mu -E_\gamma ) -(E_\gamma ^2 -\vec{k}^2_\gamma )/
2(E_\mu -E_\gamma ), \vec{\tilde k}=E_\mu \vec{k}_\gamma
/(E_\mu -E_\gamma )
\end{eqnarray}

        This formula determines the radiation probability on the set
trajectory and must be averaged in all possible trajectories.
Averaging is realized due to the distribution function which satisfies
to the standard kinetic equation
\begin{eqnarray}
\partial F/\partial t +\vec{v}\partial F/\partial \vec{r}=
nv\int \sigma (\vec{v},\vec{v}^\prime)[F(\vec{r},\vec{v},t)-
 F(\vec{r},\vec{v}^\prime,t)]d\vec{v}^\prime
\end{eqnarray}
with the initial condition
$$
F(\vec{r},\vec{r}^\prime,0,\vec{v},\vec{v}^\prime)=
\delta(\vec{r}-\vec{r}^\prime)\delta(\vec{v}-\vec{v}^\prime)
$$
        Here $\sigma (\vec{v},\vec{v}^\prime)$ is the scattering
cross-section in Couloumb field with the account of full screening
\begin{eqnarray}
\sigma (\vec{v},\vec{v}^\prime)={4E_\mu^2 Z^2 \alpha^2 \over
[(\vec{p}^\prime -\vec{p})^2+\chi^2]},
\end{eqnarray}

where $\chi $- is the screening constant in Thomas-Fermi approximation

        In small angle approximation after integration by coordinates
the kinetic equation transforms into Fokker-Planck equation
\begin{eqnarray}
\partial F/\partial \tau +i(a+b\theta^2/2)F=q\Delta F
\end{eqnarray}
where
\begin{eqnarray}
 q={2\pi nZ^2\alpha ^2\over E_\mu ^2}\ln \left({\theta _2\over \theta _1}\right),\qquad
 a={\omega _0^2\over 2E_\gamma }+{m_\mu ^2E_\gamma \over 2E_\mu (E_\mu -E_\gamma )},\qquad
 b={E_\gamma E_\mu \over E_\mu -E_\gamma },
\label{4}
\end{eqnarray}
 $\theta _1$ and $\theta _2$ are minimal and maximal angles of scattering
 in bremsstrahlung formation process, $n$ is the density of scattering centers.
The details to solve Fokker-Planck equation can be found in \cite{BKF73}.
Here we show  only the final result
\begin{eqnarray}
F(\theta ,\theta^\prime ,\tau)=\exp(\alpha(\tau )\theta^{\prime 2}+
\beta(\tau)\theta \theta^\prime +\gamma(\tau )),
\end{eqnarray}
where
$$
\alpha (\tau)=-\sqrt{ib/8q} coth (\sqrt{2ibq} \tau),
$$
$$
\beta (\tau )=\sqrt{ib/2q} sinh^{-1} (\sqrt{2ibq} \tau),
$$
$$
\gamma (\tau )=-ia\tau -ln (sinh (\sqrt{2ibq} \tau))
+\theta ^2 \alpha (\tau)+ln(\sqrt{ib/8q}).
$$
For ${\cal L}(\vec{p}^\prime ,\vec{p})$ in small angle approximation
we have
\begin{eqnarray}
 {\cal L}(\vec{p}^\prime ,\vec{p})=R_1+R_2\theta \theta ^\prime ,
\end{eqnarray}
where
 $$
 R_1={m_\mu ^2 E_\gamma ^2 \over E_\mu ^2 (E_\mu -E_\gamma )^2},\qquad
 R_2={E_\mu ^2+(E_\mu -E_\gamma )^2 \over (E_\mu -E_\gamma )^2}.
 $$

 As a result the quasiclassic approximation calculation gives the probability
 of bremsstrah\-lung in the following form:

\begin{eqnarray}
 dW={\alpha \over 2\pi } q {E_\gamma dE_\gamma \over b}
 \left [  {b\over 3a^2 } R_1G(s)+{4 \over 3a}R_2 \Phi (s) \right ]
\end{eqnarray}

 Here
 $s=a/4\sqrt{bq}$, $G(s)$ and $\Phi (s)$ are the famous Migdals functions.

        The indefinity in $\ln (\theta _2/ \theta _1)$ is the result of
transition to Fokker-Planck equation in approximation of small angles.

The argument
of logarithm contains the indefinite numerical factor. This factor can be
defined in  the
limited case $(s \gg 1)$
\footnote {Semiclassically, this condition corresponds to
$\theta _{br} \gg \theta_ s$}.

 In this case  formula (5) goes over to usual
bremsstrahlung formula Bethe-Heitler. The comparison of the limited
expression with the
formula of Petrukhin and Shestakov \cite{LKV},\cite{PS} gives us
\begin{eqnarray}
\ln \left({\theta _2\over \theta _1}\right) = \xi (s) L
\end{eqnarray}
where $L$ from \cite{LKV} is

\begin{eqnarray}
L =
 \ln \left({126m_\mu Z^{-2/3}/m_e \over
1+189\sqrt {e}\delta Z^{-1/3}/m_e }\right), & \mbox{if} \quad Z>10,
\end{eqnarray}

\begin{eqnarray}
L=\ln  \left({189m_\mu Z^{-1/3}/m_e \over
1+189\sqrt {e}\delta Z^{-1/3}/m_e }\right), & \mbox{if} \quad Z\le 10
\end{eqnarray}

Here $\delta = m_\mu ^2u/2E_\mu (1-u)$ is the minimum momentum transfer
to the nucleus and $e=2.718$.
(for the electron $L= \ln (190 Z^{-1/3})$ ).

       Therefore we can define the value $s$:
\begin{eqnarray}
s={1\over 8}\left ({1\over 2\pi}{1\over \gamma _\mu }{u\over (1-u)}
{m_\mu ^3\over nZ^2\alpha ^2L}\right )^{1/2},\quad u={E_\gamma \over E_\mu },
\qquad \gamma _\mu ={E_\mu \over m_\mu}.
\end{eqnarray}

Phenomenological factor $\xi (s)$, according to Migdal, equals:
\begin{eqnarray}
 \xi(s) = \left \{
\begin{array}{cl}
1, & \mbox{if} \quad s>1 \\
1+\ln(s)/\ln(s_1),& \mbox{if} \quad s_1<s<1 \\
2, & \mbox{if}\quad s<s_1
\end{array} \right.
\end{eqnarray}

where the value $s_1$:
\begin{eqnarray}
 s_1^{1/2} =\left \{
\begin{array}{rl}
m_eZ^{1/3}/ 189 m_{\mu}, & \mbox{if} \quad Z<10 \\
m_eZ^{2/3}/ 126 m_{\mu}  ,& \mbox{if} \quad Z>10\\
\end{array} \right.
\end{eqnarray}

        The obtained result has the logarithmic accuracy and does not contain
Coulomb corrections. The use of the more accurate formula \cite{Bug}
as a limit case is difficult because it has not only non-logarithmic terms
but also corrections proportional to  $Z$ and $1/Z$.
The accuracy of Migdal`s calculations is discussed at first in \cite{Perl}.
Recently in \cite{BK97} the calculation
with more accuracy has been performed for electron LPM effect.
These calculations showed that the Migdal calculations have the 10-15 \% accuracy.
%%%%%%%%%%%%%%%%%%%%%%%%%%%%%%%%%%%%%%%%%%%%%%%%
\subsection{The longitudinal effect}
\vspace{0.5cm}
%%%%%%%%%%%%%%%%%%%%%%%%%%%%%%%%%%%%%%%%%%%%%%%%
Now we consider the longitudinal effect, namely, the influence of
media polarization.
We can rewrite the expression of $a$ from formulae (\ref{4}):
\begin{eqnarray}
 a={\omega _0^2\over 2E_\gamma }+{m_\mu ^2E_\gamma \over 2E_\mu
(E_\mu -E_\gamma )}= \nonumber \\
= {m_\mu ^2E_\gamma \over 2E_\mu (E_\mu -E_\gamma )}\Biggl[
1+{\omega_0^2E_{\mu}(E_{\mu}-E_{\gamma})
\over E_{\gamma}^2m_{\mu}^2}\Biggr]  \\
\approx {m_\mu ^2E_\gamma \over 2E_\mu (E_\mu -E_\gamma )} \cdot
\Biggl[1+{\omega_0^2\over u^2m_\mu ^2}\Biggr].\nonumber
\end{eqnarray}

        Therefore we have the additional factor $d_\mu$:
\begin{eqnarray}
d_\mu=1+{\omega_0^2\over u^2m_\mu ^2}
\end{eqnarray}

	This factor changes the expression for bremsstrahlung cross-section
(the function $\Phi (s) \rightarrow \Phi(s)/d_\mu, G(s)\rightarrow G(s)/d_\mu^2$
and the value of $s\rightarrow d_\mu\cdot s$).

Finally the cross-section is

\begin{eqnarray}
{d\sigma \over du}=\alpha ^3 \left({2Z\over m_\mu }\right)^2{\xi (s)L\over u}
\left({1\over 3}{u^2G(s)\over d_\mu^2}
+\left({4\over 3}-{4\over 3}u+{2\over 3}u^2\right){\Phi (s)\over d_\mu}\right),
\end{eqnarray}
 parameter $s$ can be written as

\begin{eqnarray}
s= d_\mu\cdot {1\over 8}\left({E_{LPM} \over E_\mu }{u\over 1-u}\right)^{1/2},
\quad  E_{LPM}= {m_\mu ^4\over 2 \pi nZ^2 \alpha ^2L}.
\end{eqnarray}

 We illustrate these both phenomena by some numerical calculations.
Parameters of some media are collected in Table  \ref{tbl:1}.
%\newpage
%\centerline{Table 1}

        The condition $s < 1$ gives us a simple relation for the energy of
bremsstrahlung photon, where LPM effect is important:
$$
E_\gamma < {64 E_{\mu}^2\over E_{LPM}}
$$
The dielectric suppression (effect longitudinal density) is essential at the
condition $d_\mu > 2$:
$$
E_\gamma < {E_{\mu}\omega_0 \over E_{LPM}}.
$$
Thus,for example, the corresponding values $E_\gamma $ in water for muon energy
$E_\mu = 10^{15}$ $eV$ equals $E_\gamma < 25$ MeV (LPM) and $E_\gamma < 200$ MeV
(effect of longitudinal density)
%%%%%%%%%%%%%%%%%%%%%%%%%%%%%%%%%%%%%%%%%%%%%%%%%%%%%%%%%%%%%%%%%%%%%%%%%%%%%%
\vspace{0.5cm}
\subsection{Direct electron pair production by muon in a medium}
\vspace{0.5cm}
%%%%%%%%%%%%%%%%%%%%%%%%%%%%%%%%%%%%%%%%%%%%%%%%%%%%%%%%%%%%%%%%%%%%%%%%%%%%%%
         For the process of direct pair production (DPP) the double
     differential
     probability in the total energy pair and in the energy
     of one component (electron or positron) can be obtained,
     following the paper of Ternovskii \cite{Tern}.  The process DPP
     contains  the integration on the virtual photon momentum,
     therefore the answer is expressed through the integral on the
     Migdal functions. The probability of DPP has the following form:

\begin{eqnarray}
dW(E_\mu ,E_p,E_+)=\quad \quad \quad \qquad \qquad \quad \quad \quad \qquad \qquad \nonumber \\
\frac{2}{3\pi}nr_e^2Z^2\alpha ^2 \ln \left(\frac{\theta _2}{\theta _1}\right)\frac{dE_pdE_+}{E_p^2}
\left[ \left\{1+\left(1-\frac{E_p}{E_{\mu}}\right)^2\right\}\left\{A(s,x)+2\left(\frac{E_+^2}{E_p^2}+\left(1-\frac{E_+}{E_p}\right)^2\right)B(s,x)\right\}+\right. \nonumber \\
\left. \frac{E_p^2}{E_{\mu}^2}\left\{C(s,x)+2\left(\frac{E_+^2}{E_p^2}+\left(1-\frac{E_+}{E_p}\right)^2\right)D(s,x)\right\}+
8\frac{E_+}{E_p}\left(1-\frac{E_+}{E_p}\right)\left(1-\frac{E_p}{E_{\mu}}\right)E(s,x)\right],
\end{eqnarray}
where $E_p=E_+ +E_-,E_+$ and $E_- $ are the energies of component pair.

        Let us introduce the following variables:
\begin{eqnarray}
\rho ={E_+ -E_- \over E_+ +E_-},\quad v={E_+ +E_-\over E_\mu }=
{E_p\over E_\mu }.
\end{eqnarray}

        Thus, we have the following expression for the double differential
cross-section:

\begin{eqnarray}
{d^2\sigma \over d\rho dv}=
 {2\over 3\pi}r_e^2Z^2\alpha ^2
\ln \left({\theta _2\over \theta _1}\right) {1-v\over v} \cdot
\quad \quad \quad \qquad \qquad \quad \quad \qquad \qquad \\
((1+\beta )(A(s,x)+(1+\rho ^2)B(s,x))+\beta (C(s,x)+
(1+\rho ^2)D(s,x))+(1-\rho ^2)E(s,x))\nonumber
\label{22}
\end{eqnarray}

        Here we calculated the Jacobian
\begin{eqnarray}
dE_+dE_p \rightarrow E_\mu ^2 {vdvd\rho \over 2}
\end{eqnarray}

and denote
\begin{eqnarray}
\beta = {v^2\over 2(1-v)},\quad x= {m_\mu ^2\over m_e^2}
{v^2(1-\rho ^2)\over 4(1-v)},
\end{eqnarray}
\begin{eqnarray}
  s ={1\over 4}\left({E_{LPM} \over E_\mu }
  {1\over v(1-\rho ^2)}\right)^{1/2},\quad  E_{LPM}= {m_\mu ^4\over
  2 \pi nZ^2 \alpha ^2L}.
\end{eqnarray}

        The integration limits are:
\begin{eqnarray}
{2m_e\over E_{\mu }} <v <1-{3\sqrt{e} m_\mu \over 4E_\mu }Z^{1/3}
\label{29}
\end{eqnarray}
\begin{eqnarray}
0< | \rho | <\left(1-{6 m_\mu ^2\over E_\mu ^2 (1-v)} \right)\sqrt{1-2m_e/E_\mu v}
\label{30}
\end{eqnarray}

      The expressions for coefficients $A(s,x),B(s,x),C(s,x),D(s,x)$and
 $E(s,x)$
are given in Appendix A \footnote {
The error from using these
expressions gives us 0.7 \% accuracy at $E_{\mu} < 10^{22}$ $eV$ and is
less than 15 \%
accuracy at $E_{\mu} > 10^{24} eV$}.
The argument
of logarithm contains the indefinite numerical factor. This factor can be
defined in the
limited case $(s \gg 1)$. In this case  formula (7) goes over to usual
formula of  cross-section direct pair production.
The comparison of the limited expression with
formula of Petruchin and Kokoulin \cite{LKV},\cite{KP} gives us
\begin{eqnarray}
\ln ({\theta _2\over \theta _1}) = L_e  =L.
\end{eqnarray}
where $L_e$ from \cite{LKV}:
\begin{eqnarray}
L_e\approx \ln(189 Z^{-1/3}\sqrt{(1+x)(1+Y_e)})
\end{eqnarray}
Here
\begin{eqnarray}
Y_e={5-\rho ^2 +4 \beta (1+\rho ^2) \over
2(1+3\beta )\ln(3+1/x)-\rho ^2-2\beta (2-\rho ^2)}
\end{eqnarray}
For numerical estimations we can take out the minimal value
$(1+x)(1+Y_e)\sim 3.25$, therefore  $L=\ln(614Z^{-1/3})$.
The values of $E_{LPM}$ for water, standard rock and lead are listed
in Table  \ref{tbl:2}.
%\newpage
%\centerline{Table 2}

\vskip 0.2in
The role of LPM is essential for the process of DPP at

$$
s \leq 1,\quad \rm {i.e.} E_\mu \geq {E_{LPM}\over 16}{1\over v(1-\rho ^2)}.
$$

        The values of $E_{LPM}$ for bremsstrahlung and for DPP
differ from each other
only by the quantity $L$. This logarithm is the slowly changeable value
depending on the medium properties.

        The energy $E_\mu $ is larger for the process of DPP, since
the momentum of virtual photons $p_{\gamma ^*}^2 \geq 4m_e^2$ for the
production of $e^+e^-$ pairs. The LPM is more expressed in the symmetrical case,
i.e. $\rho =0$ or $E_+ \sim E_-$ for the direct pair production.

	The probability of production $\mu ^+ \mu^- $ is
\begin{eqnarray}
  dW_{ \mu^+\mu^- } \approx {m_e^2\over m_\mu ^2 }
  dW_{ e^+e^- }
\end{eqnarray}
therefore it should not be probably  taken into account.
%%%%%%%%%%%%%%%%%%%%%%%%%%%%%%%%%%%%%%%%%%%%%%%%%%%%%%%%%%%%%%%%%%%%%%%%%%%%%%
\vspace{0.5cm}
\section{Numerical calculation of the cross-section}
\vspace{0.5cm}
%%%%%%%%%%%%%%%%%%%%%%%%%%%%%%%%%%%%%%%%%%%%%%%%%%%%%%%%%%%%%%%%%%%%%%%%%%%%%%
\subsection{Bremsstrahlung process}
\vspace{0.5cm}
\subsubsection{The longitudinal density effect (Ter-Mikaelian effect)}
\vspace{0.5cm}
%%%%%%%%%%%%%%%%%%%%%%%%%%%%%%%%%%%%%%%%%%%%%%%%%%%%%%%%%%%%%%%%%%%%%%%%%%%%%%
     The longitudinal density effect ( the LD effect) exists irrespectivly
of the absence or presence of the LPM effect in the bremsstrahlung process.
Therefore, we evaluate the contribution from the LD effect to the
differential cross section. In Figures 1 to 4,
we calculate the differential cross sections for water, standard rock,
iron and lead,
respectively,where BH denotes "without the LPM effect".
In these figures, the differential cross sections of BH, (BH
and LD), LPM and (LPM+BH) are given. It is easily understood that the LD
effect suppresses much lower energy photon, compared to the LPM effect.
And it is also understood that the LD effect is more effective in higher
density materials. Further it should be noted the following:
the fact that BH+LD coincide with LPM+LD at smaller $u$ in from Figs.1
to 4 meaning that the LD effect is only effective in the region at smaller
$u$.
However the LD effect is not so effective compared to the LPM effect,
as in the same situation in the case of electron \footnote {See subsequent
paper}

     In order to examine the contribution from the LD effect over muon's
propagation, we calculate the corresponding integral cross section ( the
total cross section) due to the LD effect in Figures 5 and 6.

In Figure 5, we give the ratio of (BH and LD) to BH from $10^9$ $eV$ to $10^{24}$ $eV$.
 In the calculation for total cross sections throughout present work,
the lower bound for integration is $10^6$ $eV$. From the figure we could
realize that the LD effect is not effective up to $10^{12}$ $eV$.
The LD effect begins to be effective around $10^{12}$ $eV$ and makes
the integral cross sections decrease to 60 \% or more at $10^{24}$ $eV$
 in the absence of the LPM effect.
However, we notice that not a small apparent decrease of  the ratio
exclusively comes from infra-red divergence due to the Bethe-Heitler's
cross section which never influences over the real behavior of higher energy
muon and besides the LD effect is negligible compared to the LPM effect
at such higher energies , which are shown in Figure 6.  Comparing  Figure
5 with Figure 6, it is easy to understand that (1) the LD effect is only
effective and the LPM effect is never  effective up to around $10^{14}$ $eV$,
 (2) the competition occurs between the LD effect and the LPM effect from
$10^{14}$ $eV$ to $10^{20}$ $eV$, and (3) the LD effect is completely negligible
at the energies more than $10^{20}$ $eV$.
%%%%%%%%%%%%%%%%%%%%%%%%%%%%%%%%%%%%%%%%%%%%%%%%%%%%%%%%%%%%%%%%%%%%%%%%%%%%%%
\vspace{0.5cm}
\subsubsection{The LPM effect}
\vspace{0.5cm}
%%%%%%%%%%%%%%%%%%%%%%%%%%%%%%%%%%%%%%%%%%%%%%%%%%%%%%%%%%%%%%%%%%%%%%%%%%%%%%
     From Figures 1 to 4, we understand that the LPM effect is more
effective in higher density materials and becomes strong at more higher
primary energy of muon. The radiated photon at lower energies  is
proportional to $\sim 1/\sqrt{E_\gamma }$, while that due to BH is proportional
to $\sim 1/E_\gamma $, which is  essentially the same as
in the case of an electron.

In Figures 7 to 10,  we give mean free paths due to the combination of
 different effects in water, standard rock, iron and lead, respectively.
The mean free path of an electron due to the LPM effect increases as primary
enegry of muon increases and as the density of material is higher.  Roughly
speaking, elongations of the mean free paths are 20, 30, 70, 120
times as large as compared to the BH case in water, standard rock, iron and lead for
the primary energy muon of $10^{24}$ $eV$, respectively.
The mean free paths of muons with LPM effect are given in Figure
11, while those of the muon with both LPM and LD effects in
Figure 12.
Comparison of Figure 11 and Figure 12 shows that the mean free path of muon
with both LPM effect and the LD effect increases by about fifty percent
over $10^{12}$ eV and $10^{20}$ eV, which depends on the materials,
due to the existence of the LD effect compared to the LPM effect only.
%%%%%%%%%%%%%%%%%%%%%%%%%%%%%%%%%%%%%%%%%%%%%%%%%%%%%%%%%%%%%%%%%%%%%%%%%%%%%%
\vspace{0.5cm}
\subsection{Direct electron-positron pair production process}
\vspace{0.5cm}
%%%%%%%%%%%%%%%%%%%%%%%%%%%%%%%%%%%%%%%%%%%%%%%%%%%%%%%%%%%%%%%%%%%%%%%%%%%%%%
          As shown in the Appendix B, the LPM effect in the direct electron
pair production process could be completely negligible up to $10^{24}$ $eV$
even for lead. Therefore, we calculate the cross sections without the
LPM effect only.
In Figures 13-1 and 13-2, we give the virtual photon energy
spectrum for primary energies of  $10^{16}$ $eV$ and $10^{24}$ $eV$ for the materials of lead,
iron, standard rock and water, respectively. After the integration by the
energies of virtual photon, we obtain total cross sections due to the
direct electron pair productions. In Figure 14, we give the total cross
sections for lead, iron, standard rock and water. In Figure 15, we give the
mean free paths of direct pair electrons for the same materials as in
Figure 14.
\setcounter{figure}{14}
%\setcounter{figure}{16}

%%%%%%%%%%%%%%%%%%%%%%%%%%%%%%%%%%%%%%%%%%%%%%%%%%%%%%%%%%%%%%%%%%%%%%%%%%%%%%
\vspace{0.5cm}
\section{Effective energy loss}
\vspace{0.5cm}
%%%%%%%%%%%%%%%%%%%%%%%%%%%%%%%%%%%%%%%%%%%%%%%%%%%%%%%%%%%%%%%%%%%%%%%%%%%%%%
     Here let us define the effective energy loss of muon in the bremsstrahlung
process\cite{Sakumoto}
by examining the influence over "effective energy loss " to primary energy
of muon.
\begin{eqnarray}
\eta (E_\mu )= E_{\mu}\frac{N}{A}\int_{u_{min}}^{u_{max}} du\cdot u\cdot
\frac{d}{du}\sigma_{_{LPM(BH)}}(u,E_\mu),
\label{36}
\end{eqnarray}
where $\frac{d}{du}\sigma_{_{LPM(BH)}}(u,E_\mu)$ are the differential cross
sections  with and without LPM effect.
Here, we take $u_{min}=  10^6$ $eV$  and $u_{max}=1-m_\mu c^2/E_\mu $
.

In the effective energy loss by muon, the influence of the longitudinal
density effect could be completely neglected over $10^9$ $eV$ to $10^{24}$
 $eV$ in both the BH and the LPM cases.

\setcounter{figure}{16}
In Figures 16-1 and 16-2 the effective energy loss due to bremsstrahlung
process are given in both cases with and without the LPM effect
in water and iron, and standard rock and lead, respectively.
 To clear the LPM effect on the effective energy
loss, we give the ratio of effective energy loss with the LPM effect
over that without the LPM effect in the following.
\begin{eqnarray}
r(E_\mu )= {\int\limits_{u_{min}}^{u_{max}} du\cdot u\cdot \frac{d}{du}\sigma_{_{LPM}}
(u,E_\mu)\over \int\limits_{u_{min}}^{u_{max}} du\cdot u\cdot \frac{d}{du}\sigma_{_{BH}}
(u,E_\mu)}
\label{37}
\end{eqnarray}
In Figure 17, the ratio calculated in (\ref{36}) is given.
It is shown in the
figure that the LPM effect becomes effective over $10^{20}$ $eV$,
depending on the materials.

Also let us define the average energy transfer to the virtual photon due
to direct pair electron production process in the following way.

\begin{eqnarray}
\xi (E_\mu )= E_{\mu}\frac{N}{A}\int_{v_{min}}^{v_{max}} dv\int_{\rho_{min}}^{\rho_{max}}
 v\cdot {d^2\sigma\over d\rho dv}(v,E_\mu) d\rho,
\label{38}
\end{eqnarray}
where ${d^2\sigma\over d\rho dv}(v,E_\mu) d\rho $ is a
double differential cross section for $\rho$ and $v$ in the direct pair
production.
Here the lower and upper bounds for integration are defined in (\ref{29}),
(\ref{30}).
The calculated results are given in Figure 18.
As the LPM effect could be neglected completely in the direct electron
production process shown in the previous section, the numerical results of
(\ref{37}) are given in the BH cases only.
%%%%%%%%%%%%%%%%%%%%%%%%%%%%%%%%%%%%%%%%%%%%%%%%%%%%%%%%%%%%%%%%%%%%%%%%%%%%%%
\vspace{0.5cm}
\section{Depth intensity relation of muons at extremely high energies}
\vspace{0.5cm}
%%%%%%%%%%%%%%%%%%%%%%%%%%%%%%%%%%%%%%%%%%%%%%%%%%%%%%%%%%%%%%%%%%%%%%%%%%%%%%
        Studies on the depth intensity relation of muon at deeper depth
in both water and rock are indispensable for elucidation of physics
related to an extremely high energy muon. Depth intensity relation
of muon is determined by the processes of bremsstrahlung, direct pair production,
nuclear interaction and ionization loss. For the studies on the depth intensity
relation of muon at extremely high energies, examination of the fluctuation
effect from the bremsstrahlung process should be carefully made, because
it is a strong source of the fluctuation. The following calculations are
made to examine a fluctuation effect from bremsstrahlung. To obtain physically
meaningful results at extremely high energies in the calculation on the depth
intensity relation of muon one should consider not only bremsstrahlung
but also the direct electron-positron pair production, nuclear interaction
and ionization loss down to lower energies, for example at 1 GeV.

     To examine the degree of the fluctuation from bremsstrahlung process
at extremely high energies, we calculate the depth intensity relation of
muons , taking into account the bremsstrahlung process only and keeping
$E_{prim}/E_{cut}$ to be 1000, where $E_{prim}$ denotes the primary energy of the muon
and $E_{cut}$ denotes cut off energy of muon in the calculation and the related
quantities.

In Figures 19-1 and 19-2, we give the average value of muon with and
without the LPM effect in both standard rock and water. We give the average
energy of muon with primary energies of $10^{24} eV$ and $10^{23} eV$.  If the LPM
effect becomes effective, the cross section decreases so that the average
energy of muon becomes high. From Figures 19-1 and 19-2, we could understand
that [1] the average energy of muon in standard rock is higher than that of
water, [2] the LPM effect is more effective in $10^{24} eV$ than in $10^{23} eV$ in
both cases with and without the LPM effect.

In Figure 20-1 and Figure 20-2, survival probabilities of muon are given in
rock and water, respectively.  Conditions  given to the calculations are
 exactly the same as in Figures 19-2 and 21. If the LPM effect becomes
essential, the cross section decreases so that survival probabilities
become large.   We could understand the following from both Figures . [1]
Survival probability in the rock is larger than that in water. [2] The LPM
effect is much more effective in $10^{24} eV$ than in $10^{23} eV$, [3] it gives
much higher survival probabilities for every depth.

     In both Figures 21-1 and 21-2, range fluctuation of muon are given
in both
standard rock and water. From both Figures, we could conclude the
following: comparing the contrast of LPM 24-21 and LPM 23-20 in
Figure 21-1 with those in Figure 21-2, the LPM effect is more essential in
standard rock than that in water.

From Figure 21-2, Table 3 and Table 4, it is very interesting to note
that the average values and their standard deviations are larger in water
than in standard rock both with and without the LPM effect, as far as the
fluctuation of range distributions are concerned.

   In Figures 25-1,25-2 and 25-3, we show energy spectrum in water at the
depths $10^6 g/cm^2$, $4\cdot 10^6 g/cm^2$ and $8\cdot 10^6 g/cm^2$  for primary energies of
$10^{22} eV$, $10^{23} eV$ and $10^{24} eV$, in the case with the LPM effect and without
the LPM effect(BH), respectively. From Figure 22-1, we could easily
understand that the LPM effect is not essential almost over every depth and
it shows little effect at depth $8\cdot 10^6 g/cm^2$, because the chances to lose
energy increase at larger depths than that at shallower depth so that the
LPM effect begins to work. From Figure 23-2, we could understand that the
LPM effect begins to become effective around $10^{23} eV$  From 23-3, we
understand that the LPM effect becomes effective strongly, particularly at
the depth $8\cdot 10^9g/cm^2$.

    In Figures 23-1,23-2 and 23-3, we show the corresponding spectrum to
Figures 22-1,22-2 and 22-3 in the standard rock. From the comparison of 23-1,
23-2, and 23-3 with Figures 22-1, 22-2  and 22-3, we could understand that the LPM
effect is much effective in the standard rock than in water.
%%%%%%%%%%%%%%%%%%%%%%%%%%%%%%%%%%%%%%%%%%%%%%%%%%%%%%%%%%%%%%%%%%%%%%%%%%%%%%
\vspace{0.5cm}
\section{General Discussion}
\vspace{0.5cm}
%%%%%%%%%%%%%%%%%%%%%%%%%%%%%%%%%%%%%%%%%%%%%%%%%%%%%%%%%%%%%%%%%%%%%%%%%%%%%%
The practical purpose of the study of the muon with LPM effect for
neutrino astrophysics  is to offer theoretical means by which we could
determine the energy of astronomical sources with extremely high energies .
Such extremely high energy sources are expected to be found in the earth
through the detection of the extremely high energy muon due to muon
neutrino.  The extremely high energy muon is suffered from strong fluctuation
due to the LPM effect so that the determination of energy is so difficult.
Therefore, the detailed studies on extremely high energy muon are requested
for the purpose.  Another purpose of the study on extremely high energy
muon may be to clarify the mechanism of charm hadronproduction.

  From the results thus obtained in too much simplified way, we try to
conjecture results which correspond to the real experiments. To perform
calculations of the depth intensity relation of muon  which are really
useful for the analysis of the experimental data at extremely high energy,
if really exist, we must consider the direct electron pair production and
nuclear interaction except the bremsstrahlung, and energy spectrum of muon
at sea level and must make a calculation down to too much lower energy  which
corresponds to the real experiment compared to the present minimum energy
$10^{20} eV$, say, $10^{9} eV$.  Nevertheless, we could conjecture the
following:
the LPM effect becomes effective at extremely high energies and it appears
strong at larger depth due to  strong fluctuation effect, because
characteristics of relation between primary muon with extremely high
energy, say $10^{24} eV$ and that with still extremely high energy cut off,
say
$10^{21} eV$,  will be surely  kept at the relation between primary muon with
extremely high energy and that with too lower energy , say, $10^{9} eV$ which
corresponds to the real experimental conditions.   Further, over energy of
$10^{15} eV$,  the flux of prompt muon at sea level exceeds over that of
conventional muon which is analyzed in the usual experimental data.
Therefore, we could say that the LPM muon is appeared in the flux of prompt
muon and the depth intensity relation of muon at extremely high energy is
closely related to the analysis on the mechanism of the charm
hadroproduction.

%%%%%%%%%%%%%%%%%%%%%%%%%%%%%%%%%%%%%%%%%%%%%%%%%%%%%%%%%%%%%%%%%%%%%%%%%%%%%%
\vspace{0.5cm}
\section{Acknowledgement}
\vspace{0.5cm}
%%%%%%%%%%%%%%%%%%%%%%%%%%%%%%%%%%%%%%%%%%%%%%%%%%%%%%%%%%%%%%%%%%%%%%%%%%%%%%
We would like to thank prof. F.F.Ternovskii for the useful
discussion on the direct  pair production. Also we thank prof. M.L.Ter-Mikaelian
for the discussion about the longitudinal density effect.

\vspace{0.5cm}
\appendix
{\LARGE\bf Appendix A}
\section{ On calculation of Ternovskii functions }
\vspace{0.5cm}

        The integral representations of coefficients
 $A(s,x), B(s,x), C(s,x)$, $D(s,x), E(s,x)$ are given by

$$
A(s,x)=\int_{1+x}^\infty {(z-x-1)G(sz)dz\over z^2(z-1)^2},\quad
B(s,x)=\int_{1+x}^\infty {(z-x-1)\Phi (sz)dz\over z(z-1)^2},\eqno(A-1)
$$
$$
C(s,x)=x\int_{1+x}^\infty {G(sz)dz\over z^2(z-1)^2},\quad
D(s,x)=x\int_{1+x}^\infty {\Phi (sz)dz\over z(z-1)^2},\quad
E(s,x)= \int_{1+x}^\infty {G(sz)dz\over z^2}.
$$

    Now we write the approximation formulae for the functions
$A(s,x), B(s,x), C(s,x)$, $D(s,x), E(s,x)$.
The error from using these
expressions gives us 0.7 \% accuracy at $E_{\mu} < 10^{22}$ $eV$ and is less
than 15 \%
accuracy at $E_{\mu} > 10^{24}$ $eV$. Here we used the following
simple expression for the functions $G(s)$ and $\Phi (s)$.

$$
\Phi (s)\approx {6s\over 6s+1},\quad G(s)\approx {(6s)^2\over (6s)^2+1}
$$

$$
A(s,x)={18s^2\over 36s^2+1}\left(1+{72s^2\over 36s^2+1}x\right)
\ln {36s^2(1+x)^2+1\over 36s^2x^2}-{36s^2\over 36s^2+1}
$$

$$
+{216s^3\over 36s^2+1}\left(1+{36s^2-1\over 36s^2+1}x\right)
\left(Atan (6s(x+1)) -{\pi \over 2}\right)
$$

$$
B(s,x)={6s\over 6s+1}\left(1+{6sx\over 6s+1}\right)
\ln {6s(1+x)+1\over 6sx}-{6s\over 6s+1}\eqno(A-2)
$$

$$
C(s,x)=-{(36s^2)^2x\over (36s^2+1)^2}
\ln {36s^2(1+x)^2+1\over 36s^2x^2}+{36s^2\over 36s^2+1}
$$

$$
-{216s^3(36s^2-1)\over (36s^2+1)^2}x
\left(Atan (6s(x+1)) -{\pi \over 2}\right)
$$

$$
D(s,x)={6s\over 6s+1}
-{36s^2x\over (6s+1)^2}\ln {6s(1+x)+1\over 6sx}
$$

$$
E(s,x)=6s\left({\pi \over 2}-Atan(6s(x+1))\right)
$$

	At $s \gg 1/(1+x)$ in complete agreement with the results of
the theory neglecting effects of the medium we obtain

$$
A(s,x)=(1+2x)\ln\left(1+{1\over x}\right)-2,\quad B(s,x)=(1+x)\ln\left(1+{1\over x}\right)-1,
$$

$$
C(s,x)={1+2x \over 1+x}-2x\ln\left(1+{1\over x}\right), \eqno(A-3)
$$

$$
D(s,x)=1-x\ln\left(1+{1\over x}\right),\quad E(s,x)={1\over 1+x}.
$$
\par
\noindent
And at $s \ll 1/(1+x)$ we obtain the expressions

$$
A(s,x)=-36s^2\ln (6sx),\quad B(s,x)=-6s\ln (6sx),\quad C(s,x)=36s^2,
$$
$$
 D(s,x)=6s,\quad E(s,x)=3\pi s.\eqno(A-4)
$$

\vspace{0.5cm}
\appendix
{\LARGE\bf Appendix B}
\vspace{0.5cm}

Here we show that the LPM effect could be completely negligible
for the direct electron pair production process up to $10^{24}$ $eV$
even for lead, by using the Ternovskii functions defined in the Appendix A.
The double differential cross section for the direct electron pair
production (DPP) process is given in (\ref{22}). The LPM effect
is included in the Ternovskii functions from $A(s,x)$ to $E(s,x)$
defined (A-1).
The numerical orders of all the Ternovskii functions are the same.
Therefore we consider the degree of the LPM effect in the function
$A(s,x)$ in the Appendix A, as an example.

        $A(s,x)$ is given as follows:

$$
A(s,x)=\int_{1+x}^\infty {(z-x-1)G(sz)dz\over z^2(z-1)^2}\eqno(B-1)
$$
Here, we write down
$$
G(s)=1-\{1-G(s)\}\eqno(B-2)
$$
If we insert (B-2) into (B-1), then we obtain
$$
A(s,x)=
\int_{1+x}^\infty {(z-x-1)dz\over z^2(z-1)^2}-
\int_{1+x}^\infty {(z-x-1)[1-G(sz)]dz\over z^2(z-1)^2}\eqno(B-3)
$$
The first term of the right side (B-3) represents the term without LPM effect,
the second term in (B-3) describes the influence of the LPM effect.
At $G(sz)=1$ the second term vanishes. Therefore, we could rewrite the (B-1)
in the following way:
$$
A(s,x) = A(BH)+A(LPM),\eqno(B-4)
$$
\par
\noindent
where
$$
A(BH)=
\int_{1+x}^\infty {(z-x-1)dz\over z^2(z-1)^2},\eqno(B-5)
$$
$$
A(LPM)=-
\int_{1+x}^\infty {(z-x-1)[1-G(sz)]dz\over z^2(z-1)^2}\eqno(B-6)
$$
As $(1-G(sz))$ in (B-6) is a non-negative monotonically decreasing
function, $A(LPM)$ is a non-positive function, which decreases the cross
section as the logical consequence of the LPM effect. If the absolute
value of the $A(LPM)$ is negligible compared to $A(BH)$ , then,
it is equivalent to the neglect of the LPM effect.

        Now, let us examine the degree of the LPM effect , calculating
the $A(s,x), B(s,x), C(s,x)$, $D(s,x), E(s,x)$ for $E_\mu=10^{24}$ $eV$
in the lead. Thus, as the LPM effect is the strongest in the lead among four
kinds of materials under consideration, it is enough to consider the
case of the lead.

      In Table B-1 we give  $A(BH),A(LPM)-E(BH),E(LPM)$.
As it is clearly seen from the Table B-1,  all  the
functions $A(LPM)-E(LPM)$ are completely negligible for $A(BH)-E(BH)$,
respectively over all $v$ and $\rho$. This is the reason why we assert
that the LPM effect  could be completely negligible for the direct pair
production process up to $10^{24}$ $eV$.

In Table B-1, we give numerical values of  $A(BH)$, $A(LPM)$ for different $v$
and $\rho$ in the case of lead with $10^{24} eV$.  As you understand from these
numerical values, the LPM effect could be completely negligible even in the
lead with $10^{24} eV$.

\newpage
\vspace{0.5cm}
\appendix
{Figures and Table Captions }
\vspace{0.5cm}

\begin{description}
\item[Figure 1]  Differential cross sections for bremsstrahlung processes
in water.
LPM denotes the case with the LPM effect and LPM+LD denotes the case
with longitudinal effect (LD) in addition to the LPM effect, while BH
denotes the case without the LPM effect (the Bethe-Heitler case) and BH+LPM
effect denotes the case without the LPM effect and with the LD effect.
[a] denotes $10^{15}$eV, the primary energy of the muon and [b] to [ji]
denote $10^{16}$ to $10^{24}$eV, primary energy of the muons, respectively.

\item[Figure 2]  Differential cross sections for bremsstrahlung processes in
standard rock.     The notations in the figure are the same as in Figure 1

\item[Figure 3] Differential cross sections for bremsstrahlung processes in
iron.     The notations in the figure are the same as in Figure 1.

\item[Figure 4]  Differential cross sections for bremsstrahlung processes in lead
     The notations in the figure are the same as in Figure 1.

\item[Figure 5] Ratio of the total cross sections $\sigma_{BH+LD}$ to $\sigma_{BH}$ for
various materials.

\item[Figure 6] Ratio of the total cross sections $\sigma_{LPM+LD}$ to $\sigma_{LPM}$ for
various materials.

\item[Figure 7]  Mean free paths for bremsstrahlung processes in water
     The notations in the figure are the same as in Figure 1.

\item[Figure 8]  Mean free paths for bremsstrahlung processes in standard rock
     The notations in the figure are the same as in Figure 1.

\item[Figure 9]  Mean free paths for bremsstrahlung processes in iron.
     The notations in the figure are the same as in Figure 1.

\item[Figure 10]  Mean free paths for bremsstrahlung processes in lead.
     The notations in the figure are the same as in Figure 1.

\item[Figure 11]  Mean free paths for bremsstrahlung processes with the LPM
effect for various materials.

\item[Figure 12] Mean free paths for the bremsstrahlung process
with the LPM and the
LD effect for various materials.

\item[Figure 13-1]  Differential virtual gamma rays energy spectrum for direct
electron pair production processes for the primary energy with $10^{16}eV$

\item[Figure 13-2]   Differential virtual gamma rays energy spectrum for direct
electron pair production processes for the primary energy with $10^{24}eV$.

\item[Figure 14]  Total cross sections for direct electron pair production
process for various materials

\item[Figure 15] Mean free paths for direct electron pair production process for
various materials.

\item[Figure 16-1]   Effective energy loss  for bremsstrahlung process due to both
with and without LPM effct for the iron and water.

\item[Figure 16-2]   Effective energy loss  for bremsstrahlung process due to both
with and without LPM effct for the lead and s.r.

\item[Figure 17]   The Ratio of the effective energy loss with the LPM effect to
without the LPM effect for bremsstrahlung process for various materials

\item[Figure 18]  Effective energy loss for direct electron pair production
process for various materials.

\item[Figure 19-1] Decrease of average energy of muon in the standard rock
     Ordinate axis means average energy of muon divided the sampling number
of muons for both the cases with and without the LPM effect. LPM 24-21
means primary energy of muon with $10^{24}eV$ and the cut off energy of muon
with $10^{21}eV$ with the LPM effect, while BH 24-21 correspond to the case
without the LPM effect (Bethe-Heitler case). LPM 23-29 means primary energy
of muon with $10^{23}eV$ and the cut off  energy of muon with $10^{23}eV$ with
the LPM effect , while BH 23-20-  means the primary energy of muon  and the
cut off energy of muon with $10^{20}eV$ without the LPM effect .

\item[Figure 19-2]Decrease of average energy of muon in water.
     The date correspond to those in  Figure 19-1.

\item[Figure 20-1]  Survival probability of muon in standard rock
     LPM 24-21 and other notations are the same as in Figure 19-1.

\item[Figure 20-2]  Survival probability of muon in water.
     The date correspond to those in  Figure 19-1.

\item[Figure 21-1] Range distribution of muon in standard rock
     LPM 24-21 and other  notations are the same as in Figure 19-1

\item[Figure 21-2]  Range distribution of muon in water.
     Notations in the figure are the same as in Figure 19-1.
The area covered
 with the range distribution is normalized to unity.

\item[Figure 22-1]
Differential energy spectrum in water :
LPM 22-19 and BH 22-19 are of the same meaning as in Figure 19-1.

\item[Figure 22-2]
Differential energy spectrum in water :
LPM 23-20 and BH 23-20 are of the same meaning as in Figure 19-1.

\item[Figure 22-3]
Differential energy spectrum in water :
LPM 24-21 and BH 24-21 are of the same meaning as in Figure 19-1.

\item[Figure 23-1]
Differential energy spectrum in standard rock :
LPM 22-19 and BH 22-19 are of the same meaning as in Figure 19-1.

\item[Figure 23-2]
Differential energy spectrum in standard rock :
LPM 23-20 and BH 23-20 are of the same meaning as in Figure 19-1.

\item[Figure 23-3]
Differential energy spectrum in standard rock :
LPM 24-21 and BH 24-21 are of the same meaning as in Figure 19-1.

\item[Table 1] Some parameters for the bremsstrahlung process

\item[Table 2] Some parameters for the direct electron pair production process

\item[Table 3] Mean values and their standard deviations of range fluctuation of
muons in water.

\item[Table 4] Mean values and their standard deviations of range fluctuation of
muons in standard rock.

\item[Table B-1]  $A(BH)$ and $A(LPM)$ in the Ternovskii functions.
\end{description}

\newpage
\begin{table}
\caption{}
\label{tbl:1}
%\par
\begin{center}
 \begin{tabular}{|c|c|c|c|c|}\hline
  & Water $(H_2O)$ & St.rock & Lead $(Pb)$ & Fe\\
\hline  $Z$ & 7.23 & 11 & 82 & 26 \\
\hline  $A$ & 14.3 & 22 & 207.2 & 55.85 \\
\hline  $\rho (g\cdot cm^{-3})$ & 1.0 & 2.65 & 11.34 & 7.86 \\
\hline  $n(cm^{-3})$& $4.211\cdot 10^{22}$ &$7.254\cdot10^{22}$&$3.296\cdot10^{22}$&$8.475\cdot 10^{22}$\\
\hline  $L$ & 8.85     &  8.57      & 7.23  &7.99       \\
 \hline $E_{LPM} (eV)$   & $2.49\cdot 10^{24}$ & $6.44\cdot
10^{23}$  & $3.02\cdot 10^{22}$ &$1.06\cdot 10^{23}$ \\
\hline  $\omega_0 (eV)$ & 20.49 & 33.18 & 61.06 & 55.14 \\
\hline
\end{tabular}
\end{center}
\end{table}

\begin{table}
\caption{}
\label{tbl:2}
%\par
\begin{center}
 \begin{tabular}{|c|c|c|c|c|}\hline
  & Water $(H_2O)$ & St.rock & Lead $(Pb)$&Iron $(Fe)$\\
\hline $L$ & 5.76    & 5.62     & 4.95 & 5.33     \\
 \hline $E_{LPM} (eV)$   & $3.84\cdot 10^{24}$ & $0.98\cdot
10^{24}$  & $4.41\cdot 10^{22}$ & $1.59\cdot 10^{23}$  \\ \hline
\end{tabular}
\end{center}
\end{table}
\newpage

\begin{table}
\caption{}
\label{tbl:3}
%\par
\begin{center}
 \begin{tabular}{|cc|cc|cc|}\hline
\multicolumn{2}{|c|}{Muon energy} & \multicolumn{2}{c|}{BH}  &
\multicolumn{2}{c|}{LPM}  \\
\hline    $E_{prim} [eV]$  & $E_{cut} [eV]$ & $<R>$ & $\sigma$ & $<R>$ &
$\sigma $ \\
\hline  $10^{24}$ & $10^{21}$ & 34.28  & 14.60 & 38.44  & 16.93 \\
\hline  $10^{23}$ & $10^{20}$ & 32.71  & 14.53 & 33.79  & 15.00 \\
\hline  $10^{22}$ & $10^{19}$ & 33.85  & 14.97 & 33.95  & 15.04 \\
\hline  $10^{21}$ & $10^{18}$ & 35.00  & 15.39 & 34.99  & 15.42 \\
\hline  $10^{20}$ & $10^{17}$ & 36.09  & 15.77 & 36.07  & 15.80 \\
\hline
\end{tabular}
\end{center}
\end{table}

\begin{table}
\caption{}
\label{tbl:4}
%\par
\begin{center}
 \begin{tabular}{|cc|cc|cc|}\hline
\multicolumn{2}{|c|}{Muon energy} & \multicolumn{2}{c|}{BH}  &
\multicolumn{2}{c|}{LPM}  \\
\hline    $E_{prim} [eV]$  & $E_{cut} [eV]$ & $<R>$ & $\sigma$ & $<R>$ &
$\sigma$\\
\hline  $10^{24}$ & $10^{21}$ & 26.36  & 11.23 & 35.30  & 15.82 \\
\hline  $10^{23}$ & $10^{20}$ & 25.15  & 11.17 & 27.32  & 12.09 \\
\hline  $10^{22}$ & $10^{19}$ & 26.03  & 11.51 & 26.31  & 11.66 \\
\hline  $10^{21}$ & $10^{18}$ & 26.91  & 11.84 & 26.91  & 11.86 \\
\hline  $10^{20}$ & $10^{17}$ & 27.75  & 12.13 & 27.74  & 12.15 \\
\hline
\end{tabular}
\end{center}
\end{table}
%%%%%%%%%%%%%%%%%%%%%%%%%%%%%%%%%%%%% table  5 %%%%%%%%%%%%%%%%%%%%%%
\begin{table}
\caption{}
\label{tbl:5}
%\par
\begin{center}
 \begin{tabular}{|l|l|l|r|}\hline
$v$                &    $\rho$  &       A (BH)       &       A (LPM)      \\
\hline
$       1.022\cdot 10^{-18}     $       &       $       0.000\cdot 10^0 $       &       $       7.157263\cdot 10        $       &       $
0.000000\cdot 10^0	$	\\
$       1.000\cdot 10^{-17}     $       &       $       0.000\cdot 10^0 $       &       $       6.701101\cdot 10        $       &       $
0.000000\cdot 10^0	$	\\
$       1.000\cdot 10^{-17}     $       &       $       5.000\cdot 10^{-1}      $       &       $       6.729869\cdot 10        $
&	$	0.000000\cdot 10^0	$	\\
$       1.000\cdot 10^{-17}     $       &       $       9.475\cdot 10^{-1}      $       &       $       6.929182\cdot 10        $
&	$	0.000000\cdot 10^0	$	\\
$       1.000\cdot 10^{-16}     $       &       $       0.000\cdot 10^0 $       &       $       6.240584\cdot 10        $       &       $
0.000000\cdot 10^0	$	\\
$       1.000\cdot 10^{-16}     $       &       $       5.000\cdot 10^{-1}      $       &       $       6.269352\cdot 10        $
&	$	0.000000\cdot 10^0	$	\\
$       1.000\cdot 10^{-16}     $       &       $       9.949\cdot 10^{-1}      $       &       $       6.698923\cdot 10        $
&	$	0.000000\cdot 10^0	$	\\
$       1.000\cdot 10^{-15}     $       &       $       0.000\cdot 10^0 $       &       $       5.780067\cdot 10        $       &       $
0.000000\cdot 10^0	$	\\
$       1.000\cdot 10^{-15}     $       &       $       5.000\cdot 10^{-1}      $       &       $       5.808835\cdot 10        $
&	$	0.000000\cdot 10^0	$	\\
$       1.000\cdot 10^{-15}     $       &       $       9.995\cdot 10^{-1}      $       &       $       6.468665\cdot 10        $
&	$	0.000000\cdot 10^0	$	\\
$       1.000\cdot 10^{-14}     $       &       $       0.000\cdot 10^0 $       &       $       5.319550\cdot 10        $       &       $
0.000000\cdot 10^0	$	\\
$       1.000\cdot 10^{-14}     $       &       $       5.000\cdot 10^{-1}      $       &       $       5.348318\cdot 10        $
&	$	0.000000\cdot 10^0	$	\\
$       1.000\cdot 10^{-14}     $       &       $       9.999\cdot 10^{-1}      $       &       $       6.238406\cdot 10        $
&	$	0.000000\cdot 10^0	$	\\
$       1.000\cdot 10^{-13}     $       &       $       0.000\cdot 10^0 $       &       $       4.859033\cdot 10        $       &       $
0.000000\cdot 10^0	$	\\
$       1.000\cdot 10^{-13}     $       &       $       5.000\cdot 10^{-1}      $       &       $       4.887801\cdot 10        $
&	$	0.000000\cdot 10^0	$	\\
$       1.000\cdot 10^{-13}     $       &       $       1.000\cdot 10^0 $       &       $       6.008148\cdot 10        $       &       $
0.000000\cdot 10^0	$	\\
$       1.000\cdot 10^{-12}     $       &       $       0.000\cdot 10^0 $       &       $       4.398515\cdot 10        $       &       $
0.000000\cdot 10^0	$	\\
$       1.000\cdot 10^{-12}     $       &       $       5.000\cdot 10^{-1}      $       &       $       4.427284\cdot 10        $
&	$	0.000000\cdot 10^0	$	\\
$       1.000\cdot 10^{-12}     $       &       $       1.000\cdot 10^0 $       &       $       5.777889\cdot 10        $       &       $
0.000000\cdot 10^0	$	\\
$       1.000\cdot 10^{-11}     $       &       $       0.000\cdot 10^0 $       &       $       3.937998\cdot 10        $       &       $
0.000000\cdot 10^0	$	\\
$       1.000\cdot 10^{-11}     $       &       $       5.000\cdot 10^{-1}      $       &       $       3.966767\cdot 10        $
&	$	0.000000\cdot 10^0	$	\\
$       1.000\cdot 10^{-11}     $       &       $       1.000\cdot 10^0 $       &       $       5.547631\cdot 10        $       &       $
0.000000\cdot 10^0	$	\\
$       1.000\cdot 10^{-10}     $       &       $       0.000\cdot 10^0 $       &       $       3.477481\cdot 10        $       &       $
0.000000\cdot 10^0	$	\\
$       1.000\cdot 10^{-10}     $       &       $       5.000\cdot 10^{-1}      $       &       $       3.506250\cdot 10        $
&	$	0.000000\cdot 10^0	$	\\
$       1.000\cdot 10^{-10}     $       &       $       1.000\cdot 10^0 $       &       $       5.317372\cdot 10        $       &       $
0.000000\cdot 10^0	$	\\
$       1.000\cdot 10^{-9}      $       &       $       0.000\cdot 10^0 $       &       $       3.016964\cdot 10        $       &       $
0.000000\cdot 10^0	$	\\
$       1.000\cdot 10^{-9}      $       &       $       5.000\cdot 10^{-1}      $       &       $       3.045733\cdot 10        $       &

$	0.000000\cdot 10^0	$	\\
$       1.000\cdot 10^{-9}      $       &       $       1.000\cdot 10^0 $       &       $       5.087114\cdot 10        $       &       $
0.000000\cdot 10^0	$	\\
$       1.000\cdot 10^{-8}      $       &       $       0.000\cdot 10^0 $       &       $       2.556447\cdot 10        $       &       $
0.000000\cdot 10^0	$	\\
$       1.000\cdot 10^{-8}      $       &       $       5.000\cdot 10^{-1}      $       &       $       2.585216\cdot 10        $       &
$	0.000000\cdot 10^0	$	\\
$       1.000\cdot 10^{-8}      $       &       $       1.000\cdot 10^0 $       &       $       4.856855\cdot 10        $       &       $
0.000000\cdot 10^0	$	\\
$       1.000\cdot 10^{-7}      $       &       $       0.000\cdot 10^0 $       &       $       2.095930\cdot 10        $       &       $
0.000000\cdot 10^0	$	\\
$       1.000\cdot 10^{-7}      $       &       $       5.000\cdot 10^{-1}      $       &       $       2.124699\cdot 10        $       &
$	0.000000\cdot 10^0	$	\\
$       1.000\cdot 10^{-7}      $       &       $       1.000\cdot 10^0 $       &       $       4.626597\cdot 10        $       &       $
0.000000\cdot 10^0	$	\\
$       1.000\cdot 10^{-6}      $       &       $       0.000\cdot 10^0 $       &       $       1.635413\cdot 10        $       &       $
0.000000\cdot 10^0	$	\\
$       1.000\cdot 10^{-6}      $       &       $       5.000\cdot 10^{-1}      $       &       $       1.664182\cdot 10        $       &
$	0.000000\cdot 10^0	$	\\
$       1.000\cdot 10^{-6}      $       &       $       1.000\cdot 10^0 $       &       $       4.396338\cdot 10        $       &       $
0.000000\cdot 10^0	$	\\
$       1.000\cdot 10^{-5}      $       &       $       0.000\cdot 10^0 $       &       $       1.174898\cdot 10        $       &       $
-3.030322\cdot 10^{-6}	$	\\
$       1.000\cdot 10^{-5}      $       &       $       5.000\cdot 10^{-1}      $       &       $       1.203666\cdot 10        $       &
$	-1.751002\cdot 10^{-6}	$	\\
$       1.000\cdot 10^{-5}      $       &       $       1.000\cdot 10^0 $       &       $       4.166079\cdot 10        $       &       $
0.000000\cdot 10^0	$	\\
$       1.000\cdot 10^{-4}      $       &       $       0.000\cdot 10^0 $       &       $       7.145755\cdot 10^0      $       &
$	-1.698723\cdot 10^{-4}	$	\\
$       1.000\cdot 10^{-4}      $       &       $       5.000\cdot 10^{-1}      $       &       $       7.432968\cdot 10^0      $
&	$	-1.002189\cdot 10^{-4}	$	\\
$       1.000\cdot 10^{-4}      $       &       $       1.000\cdot 10^0 $       &       $       3.935811\cdot 10        $       &       $
0.000000\cdot 10^0	$	\\
$       1.000\cdot 10^{3}       $       &       $       0.000\cdot 10^0 $       &       $       2.645587\cdot 10^0      $       &       $
-4.427908\cdot 10^{3}	$	\\
$       1.000\cdot 10^{3}       $       &       $       5.000\cdot 10^{-1}      $       &       $       2.910863\cdot 10^0      $
&	$	-2.918965\cdot 10^{3}	$	\\
$       1.000\cdot 10^{3}       $       &       $       1.000\cdot 10^0 $       &       $       3.705456\cdot 10        $       &       $
0.000000\cdot 10^0	$	\\
$       1.000\cdot 10^{2}       $       &       $       0.000\cdot 10^0 $       &       $       7.112332\cdot 10^{2}    $       &
$	-2.047297\cdot 10^{-4}	$	\\
$       1.000\cdot 10^{2}       $       &       $       5.000\cdot 10^{-1}      $       &       $       1.066582\cdot 10^{-1}
$	&	$	-3.217814\cdot 10^{-4}	$	\\
$       1.000\cdot 10^{2}       $       &       $       1.000\cdot 10^0 $       &       $       3.474261\cdot 10        $       &       $
0.000000\cdot 10^0	$	\\
$       1.000\cdot 10^{-1}      $       &       $       0.000\cdot 10^0 $       &       $       1.171890\cdot 10^{-5}   $
&	$	-2.204098\cdot 10^{-13}	$	\\
$       1.000\cdot 10^{-1}      $       &       $       5.000\cdot 10^{-1}      $       &       $       2.077567\cdot
10^{-5}	$	&	$	-7.148156\cdot 10^{-13}	$	\\
$       1.000\cdot 10^{-1}      $       &       $       1.000\cdot 10^0 $       &       $       3.233732\cdot 10        $       &       $
0.000000\cdot 10^0	$	\\
$       1.000\cdot 10^0 $       &       $       0.000\cdot 10^0 $       &       $       0.000000\cdot 10^0      $       &       $
0.000000\cdot 10^0	$	\\
$       1.000\cdot 10^0 $       &       $       5.000\cdot 10^{-1}      $       &       $       0.000000\cdot 10^0      $       &
$	0.000000\cdot 10^0	$	\\
$       1.000\cdot 10^0 $       &       $       1.000\cdot 10^0 $       &       $       2.363193\cdot 10^{-10}  $       &
$	0.000000\cdot 10^0	$	\\

\hline
\end{tabular}
\end{center}
\end{table}

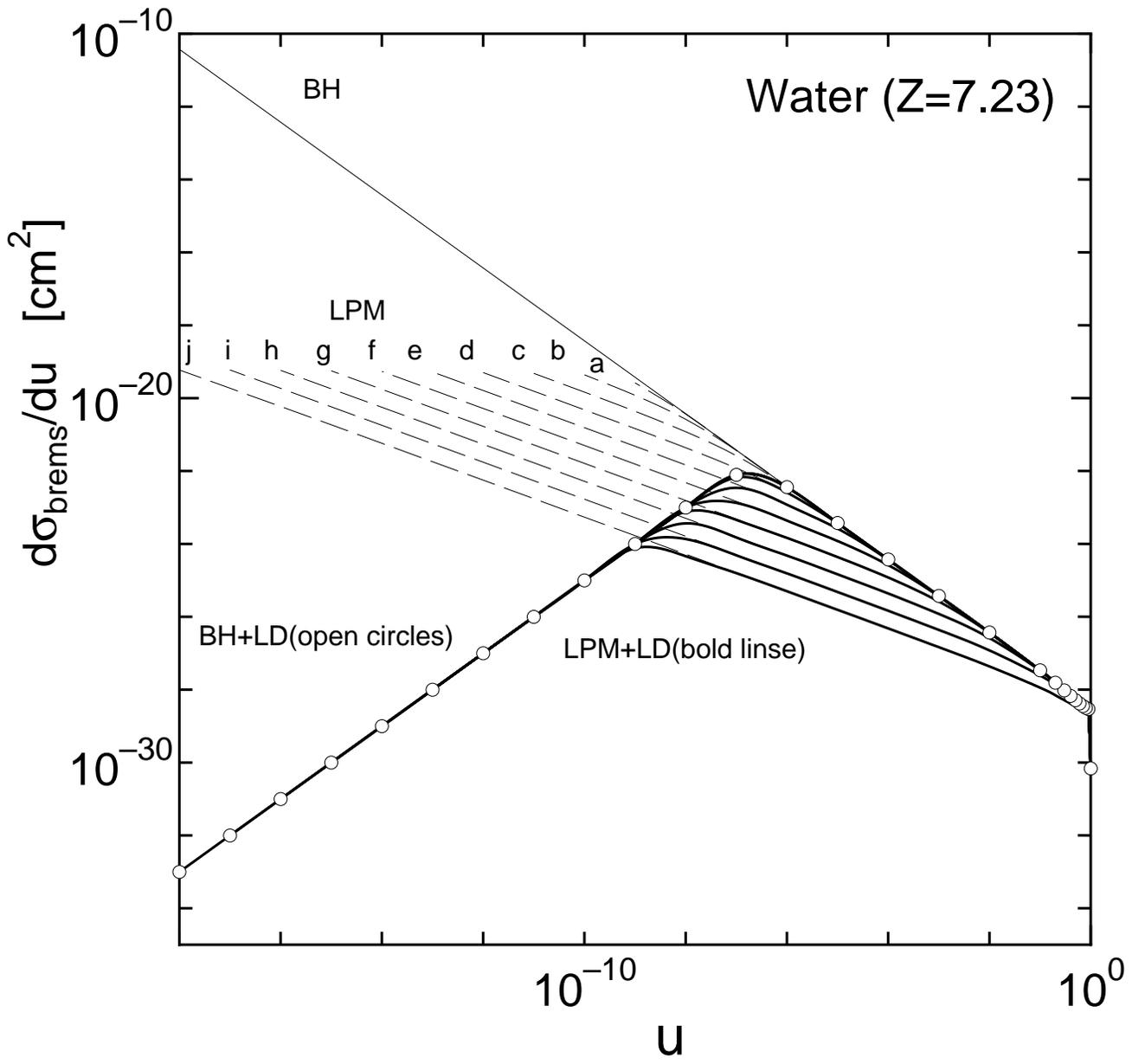
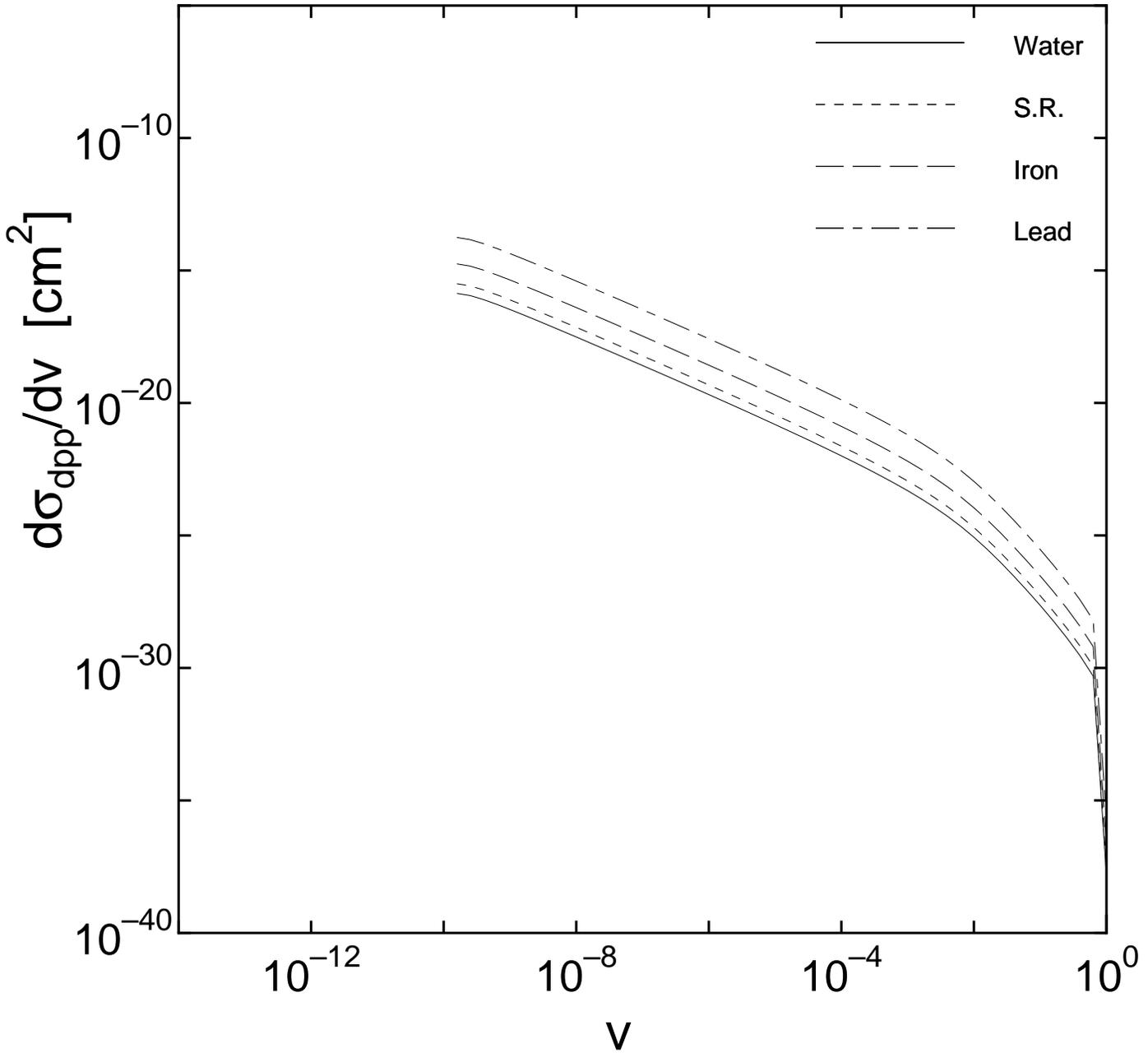
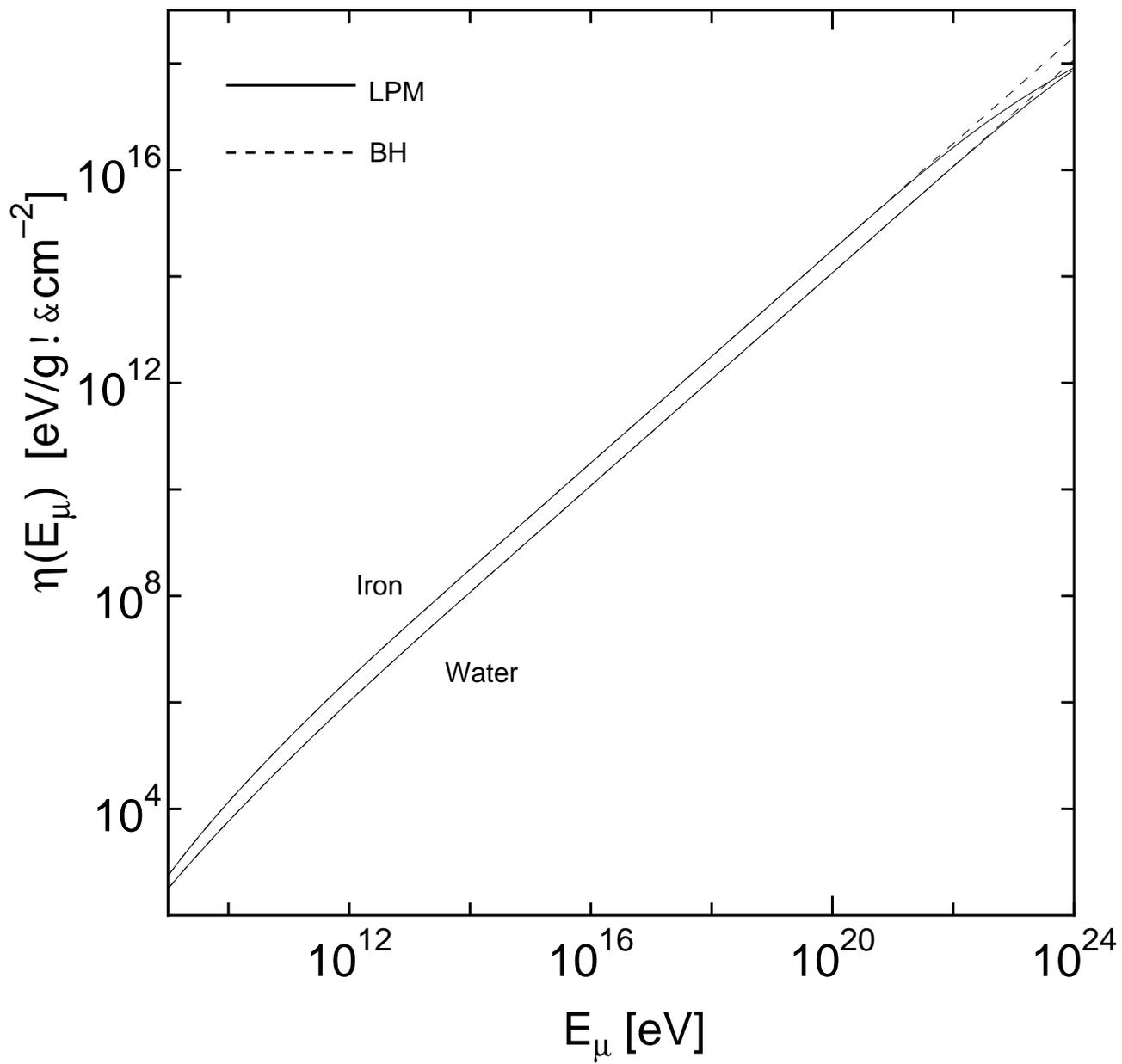
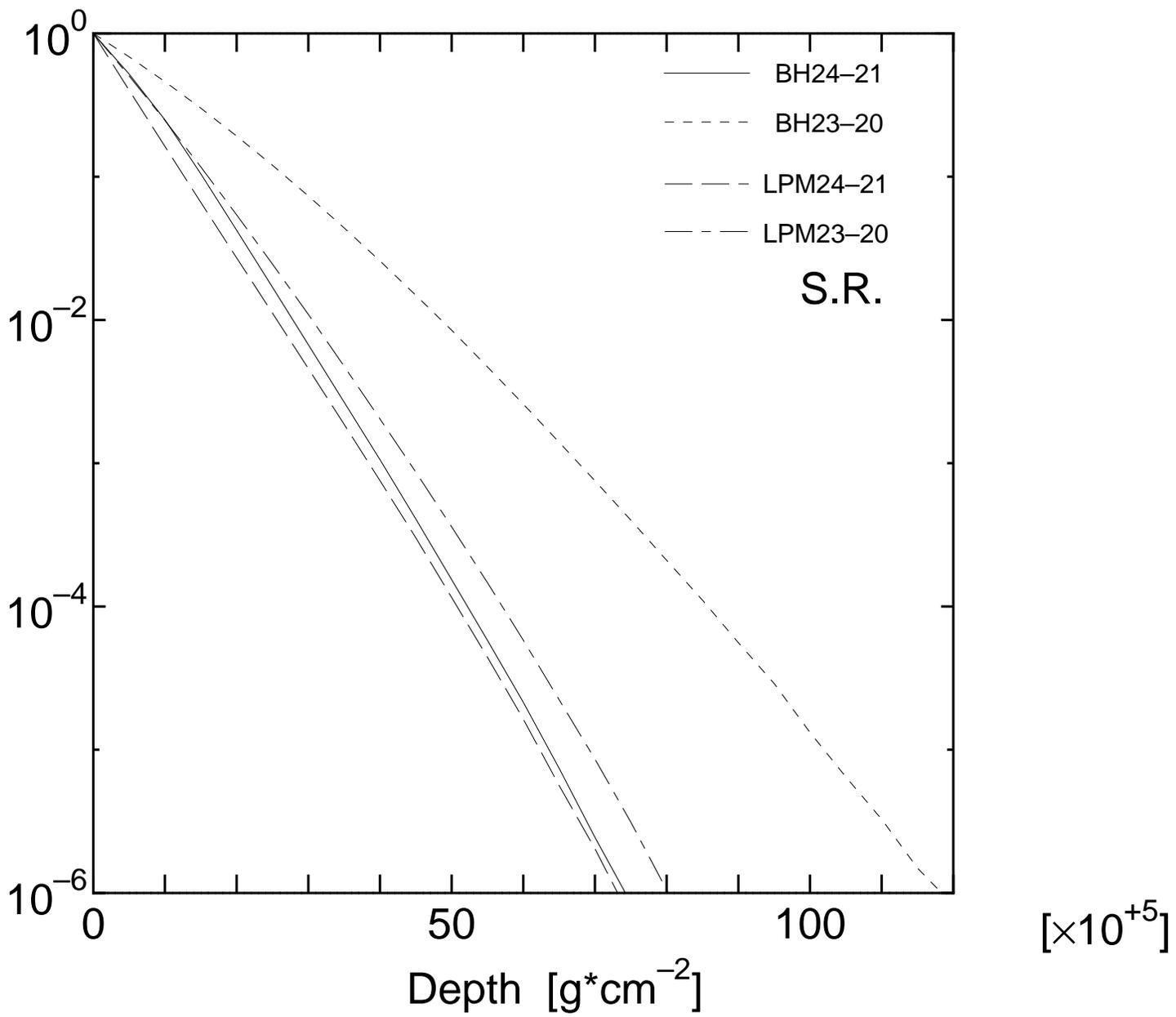
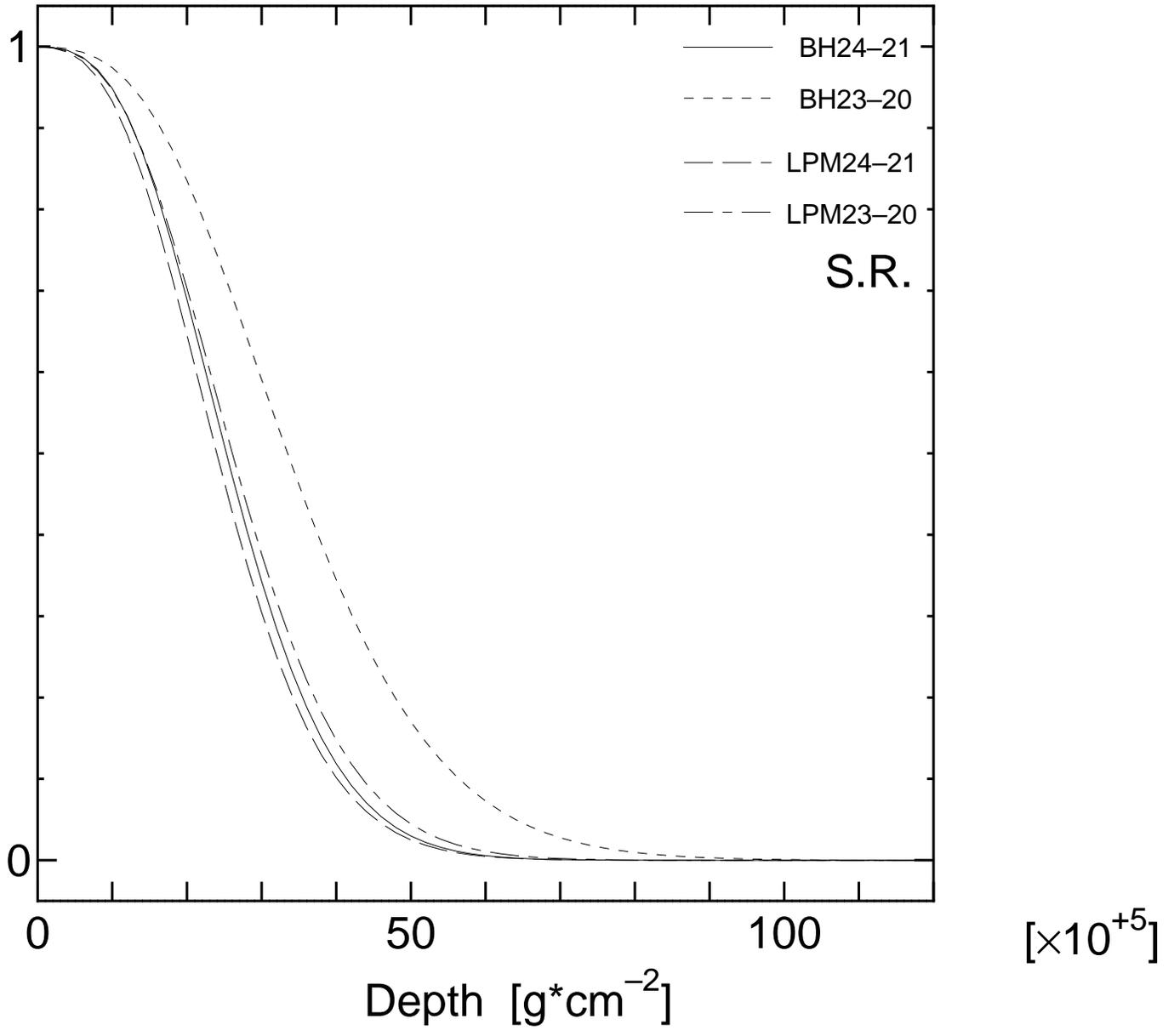
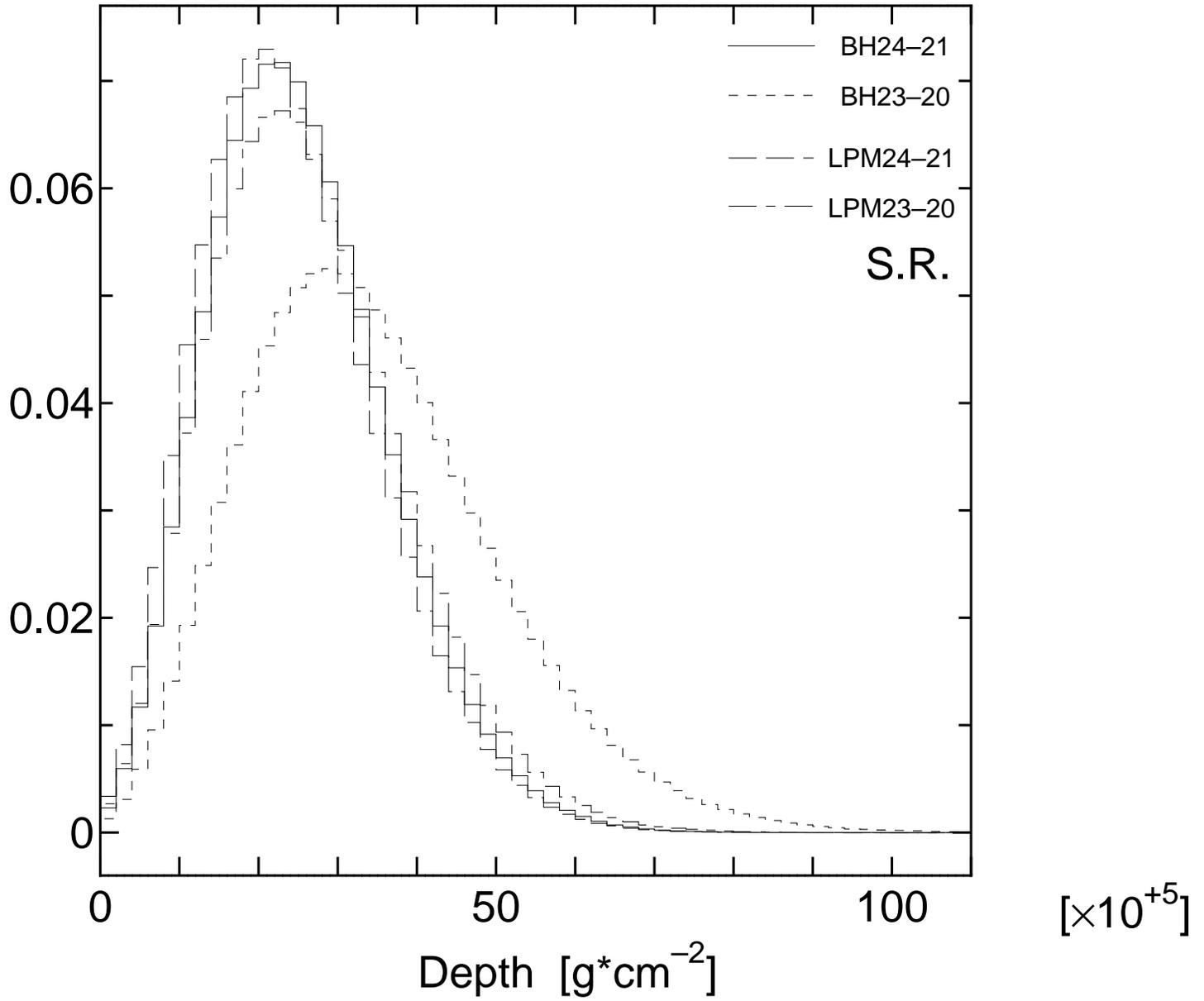
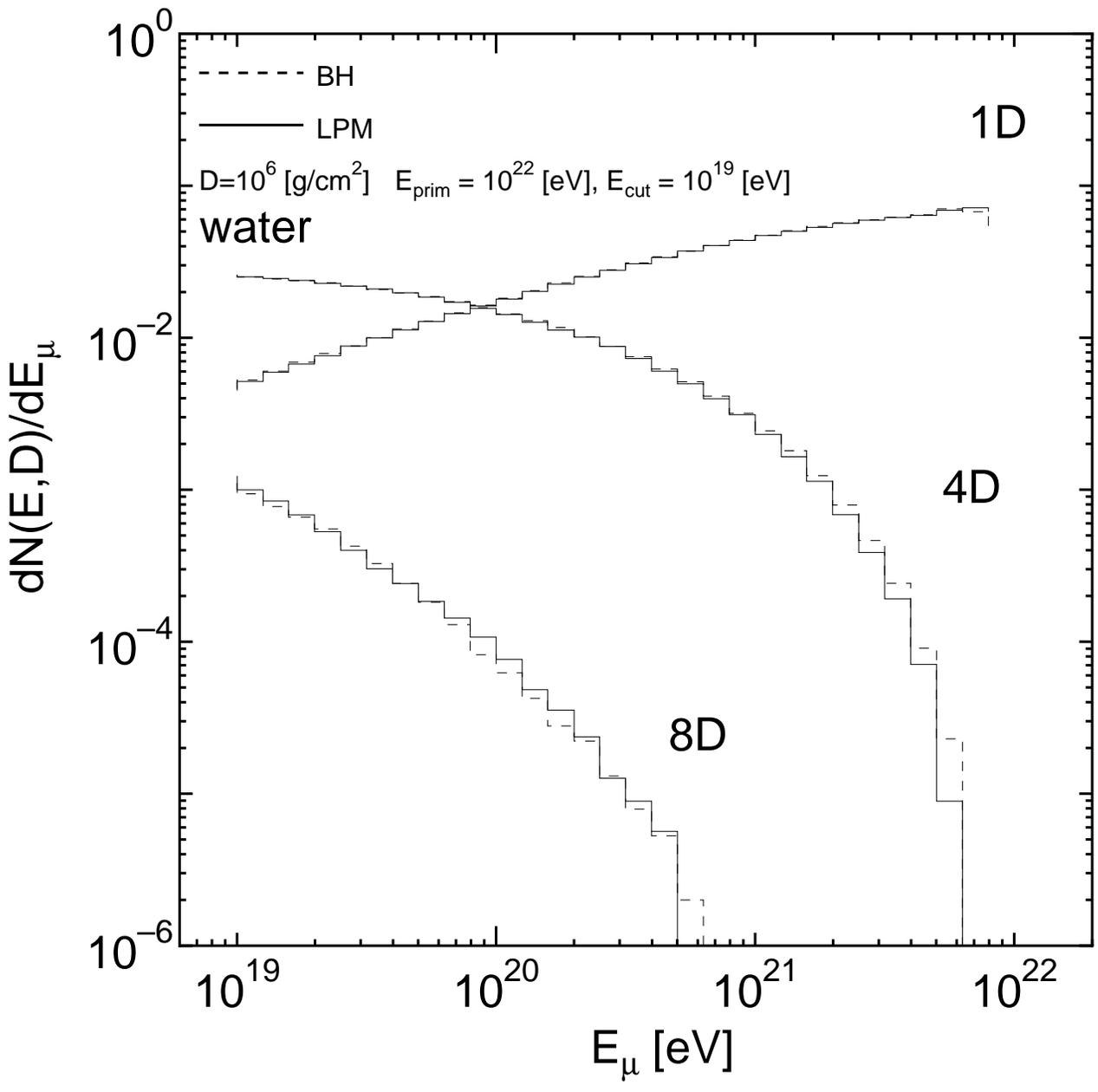
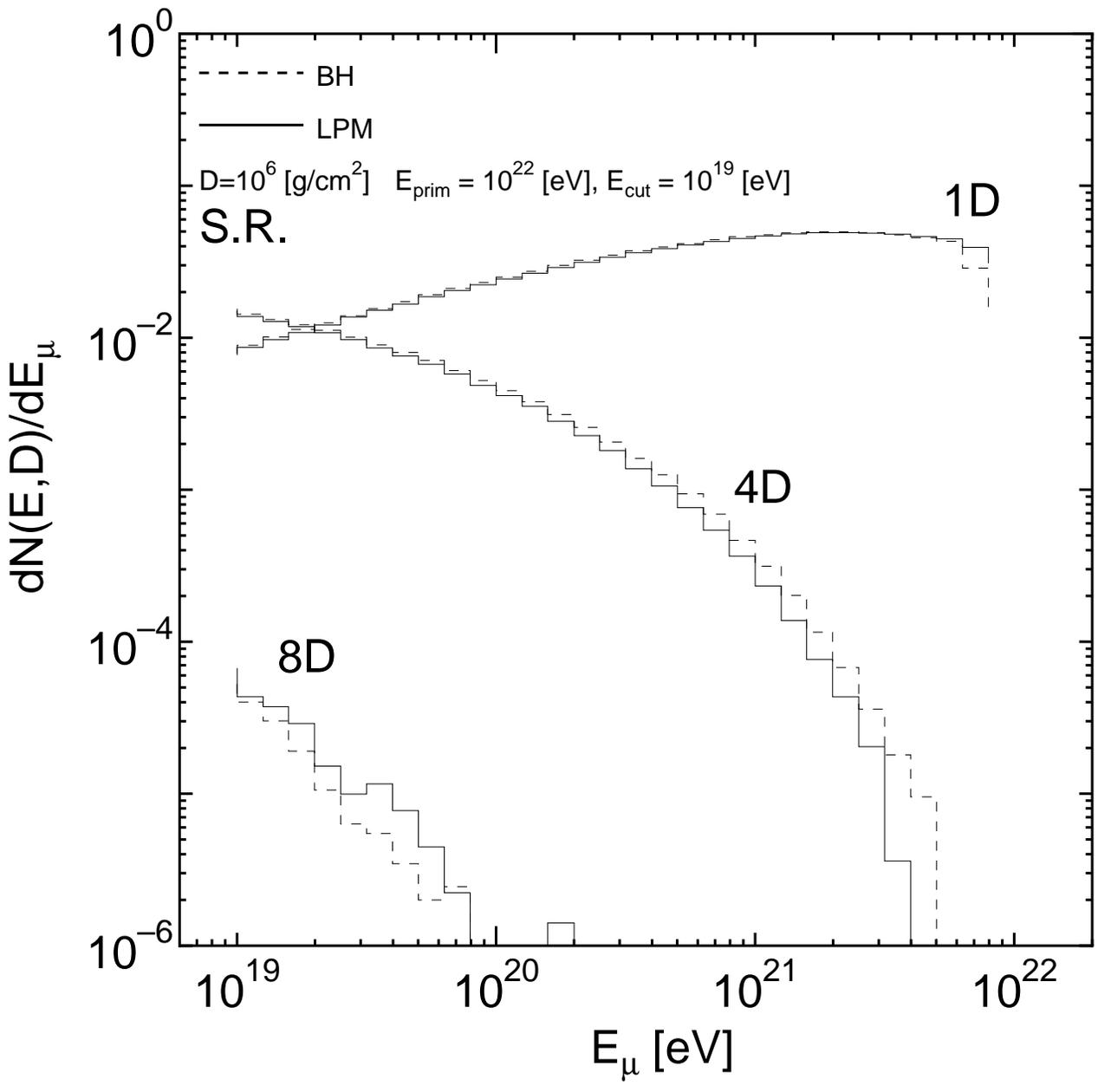
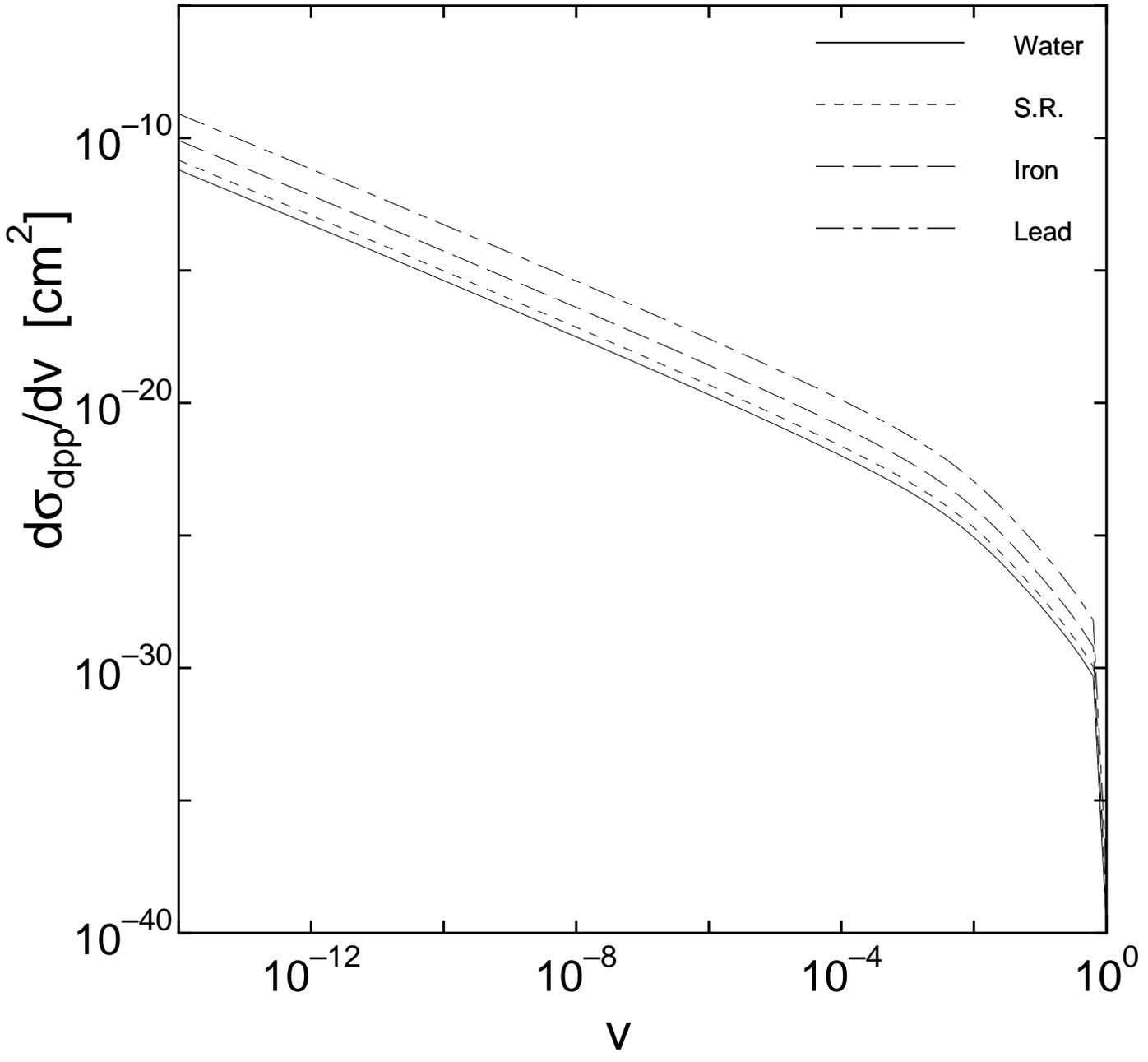
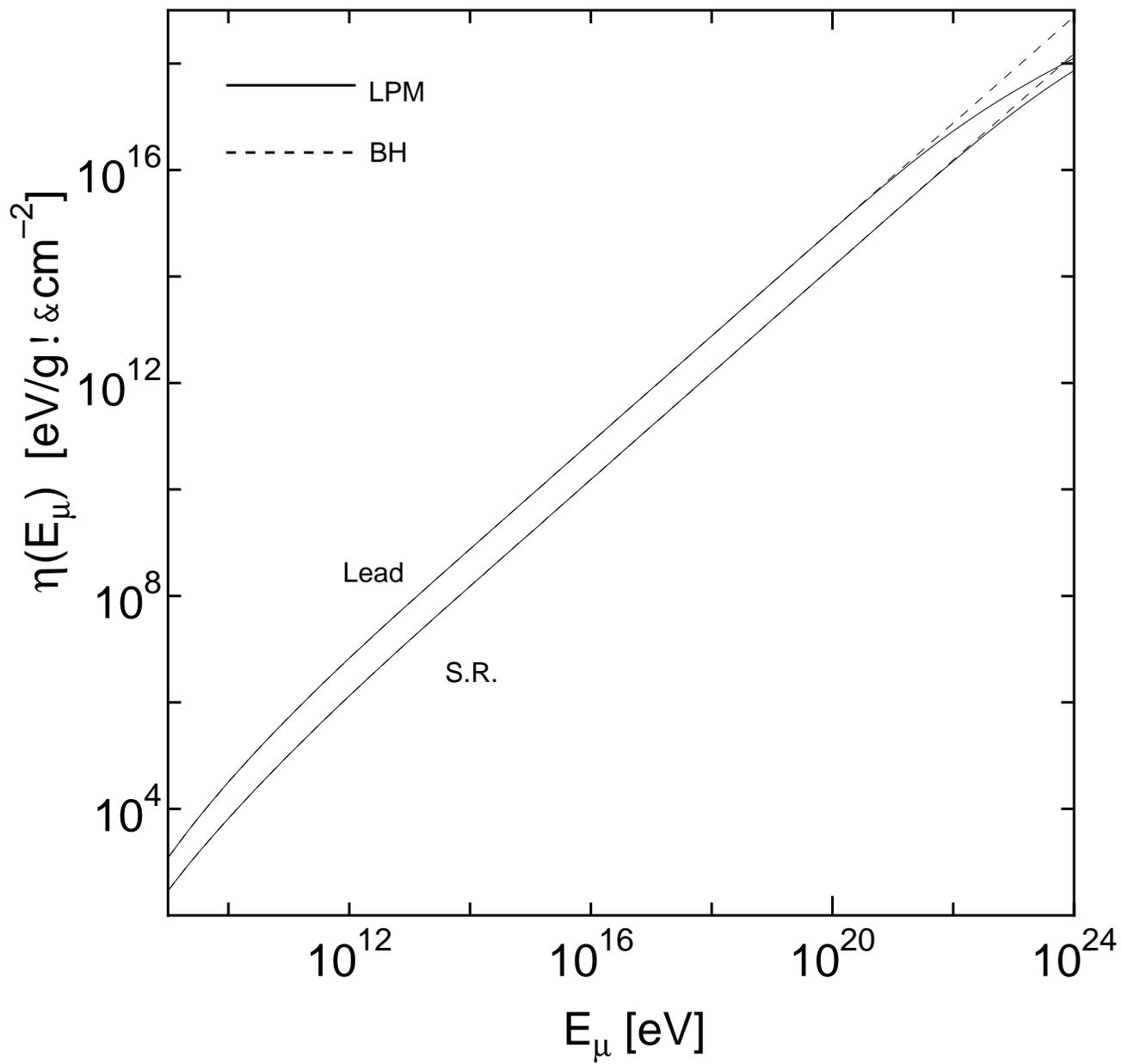
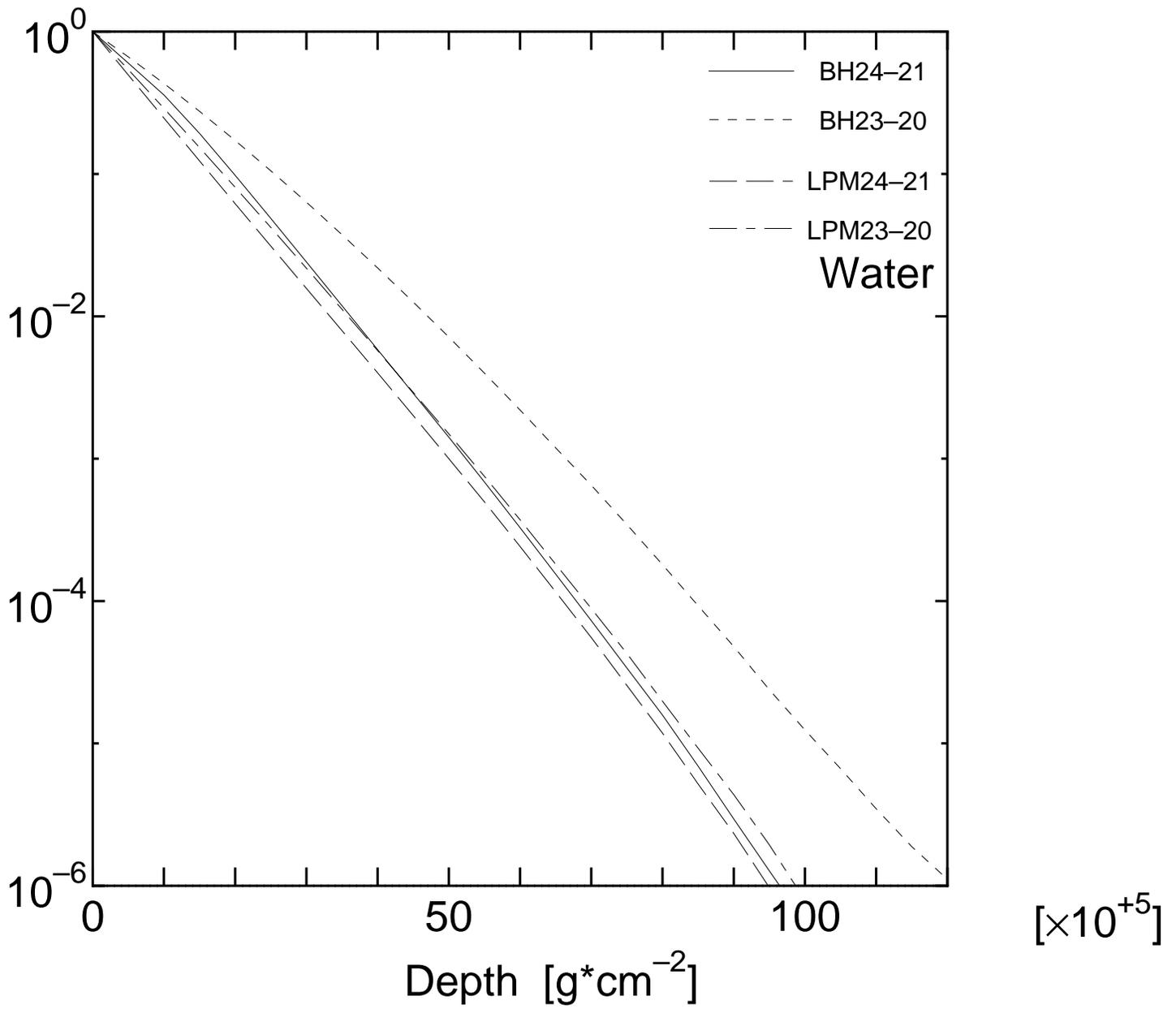
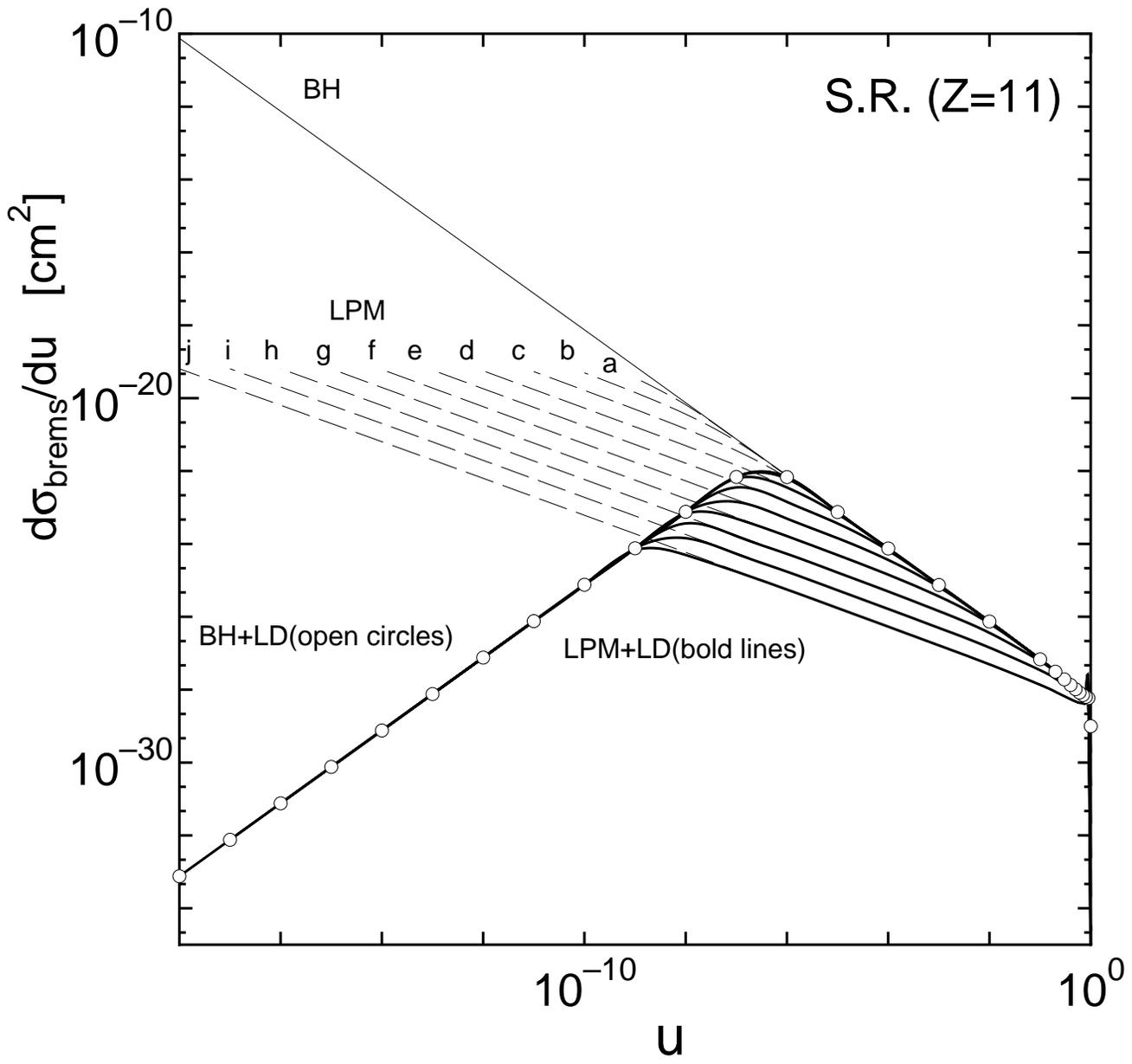
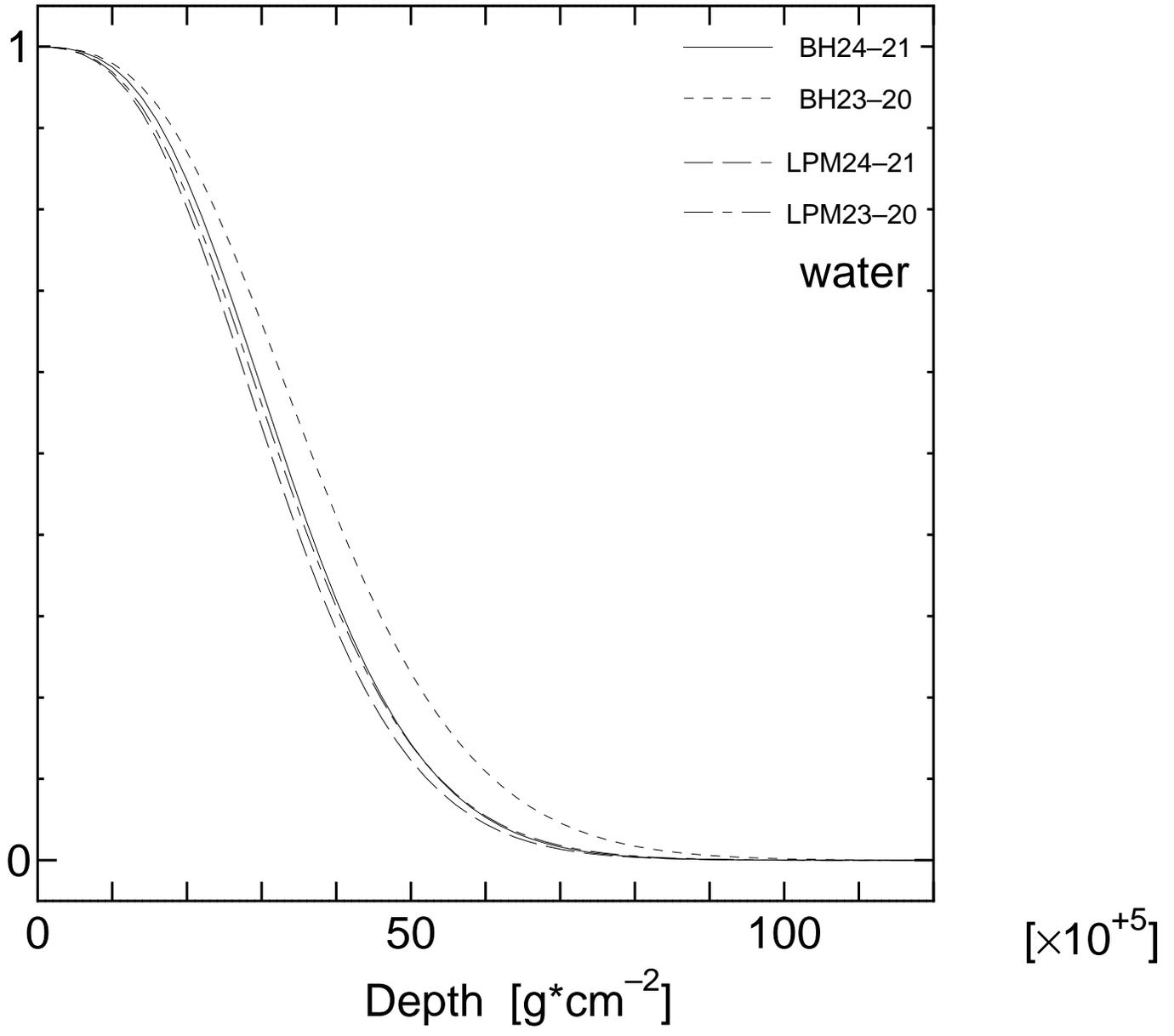
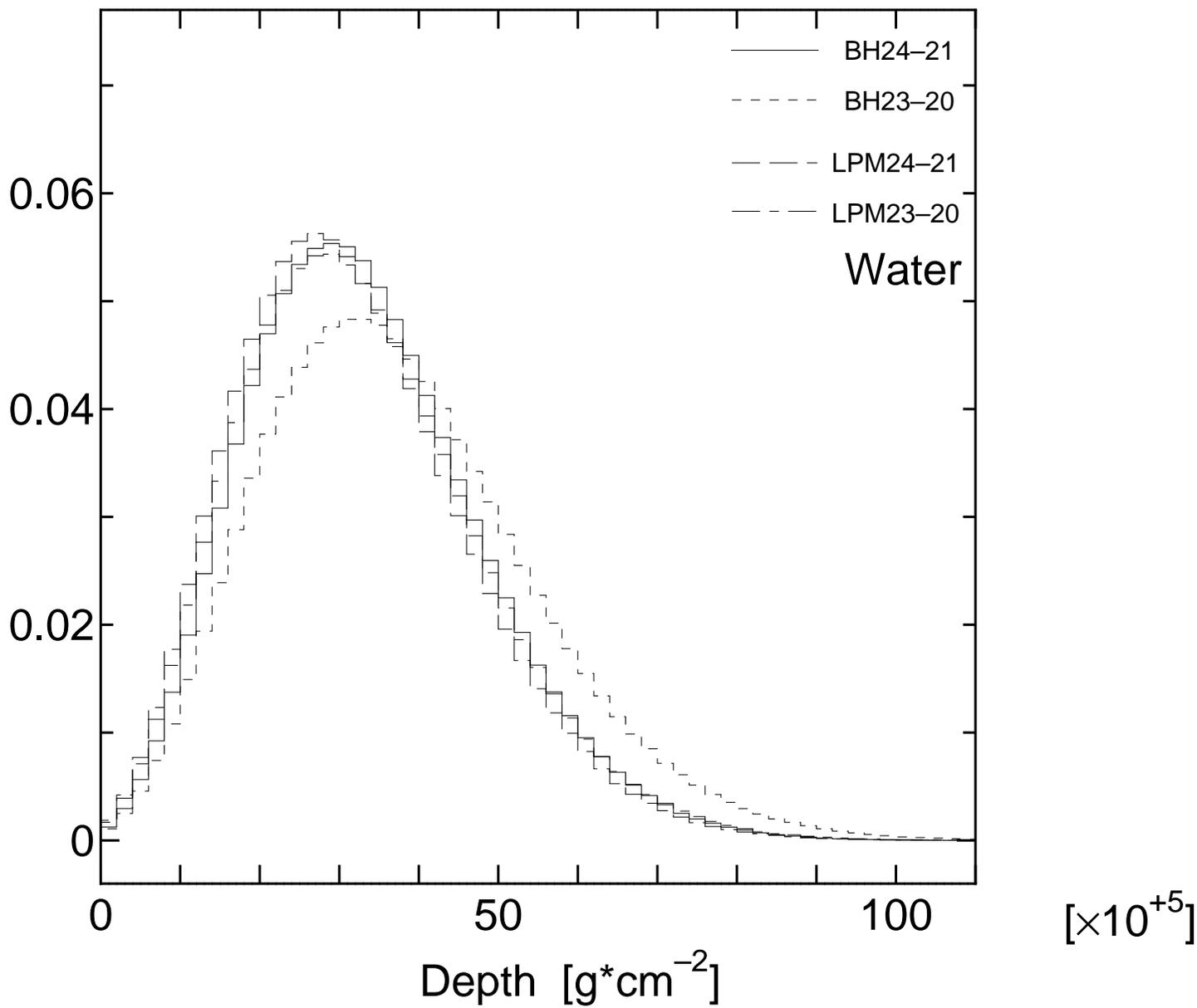
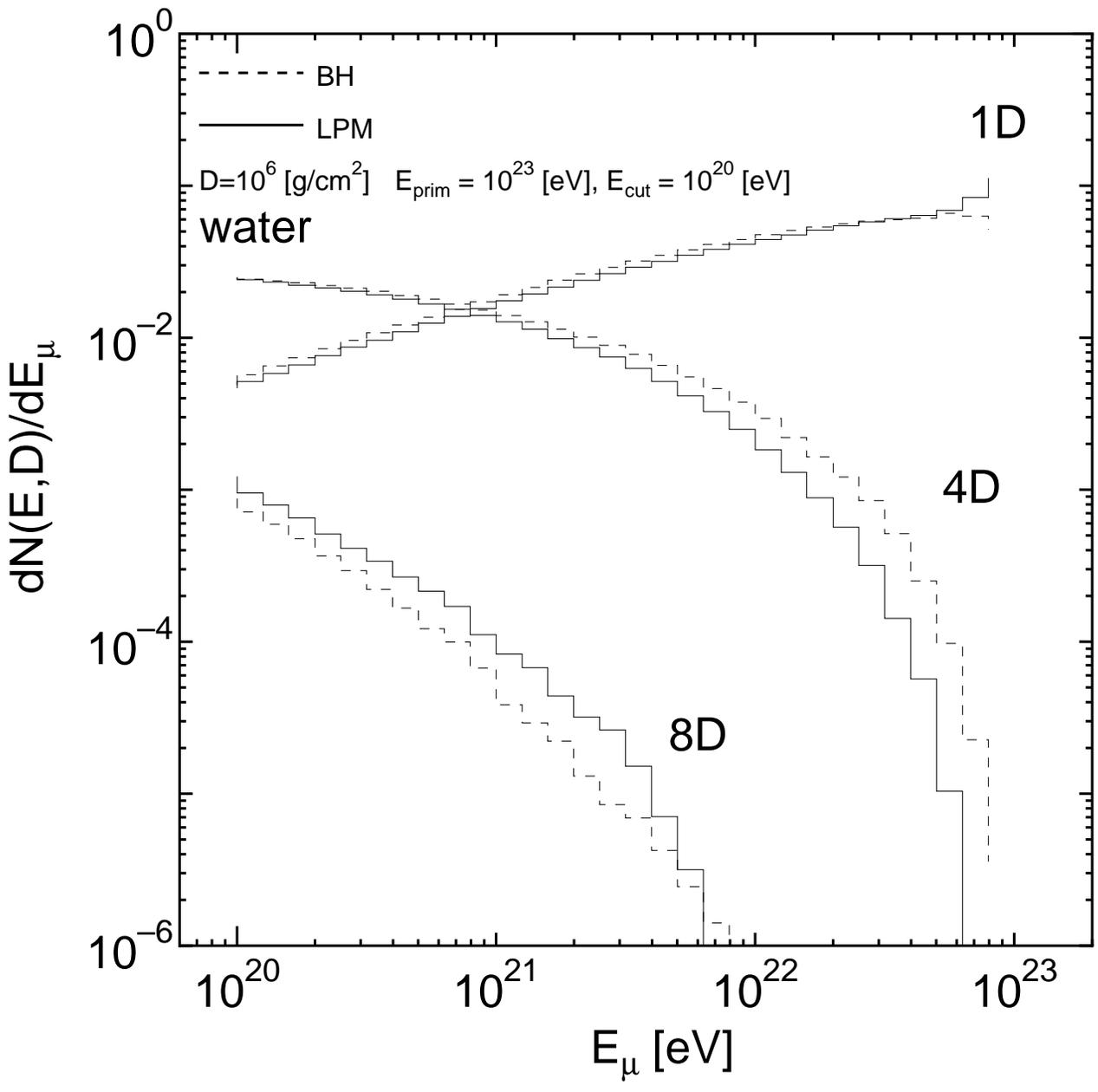
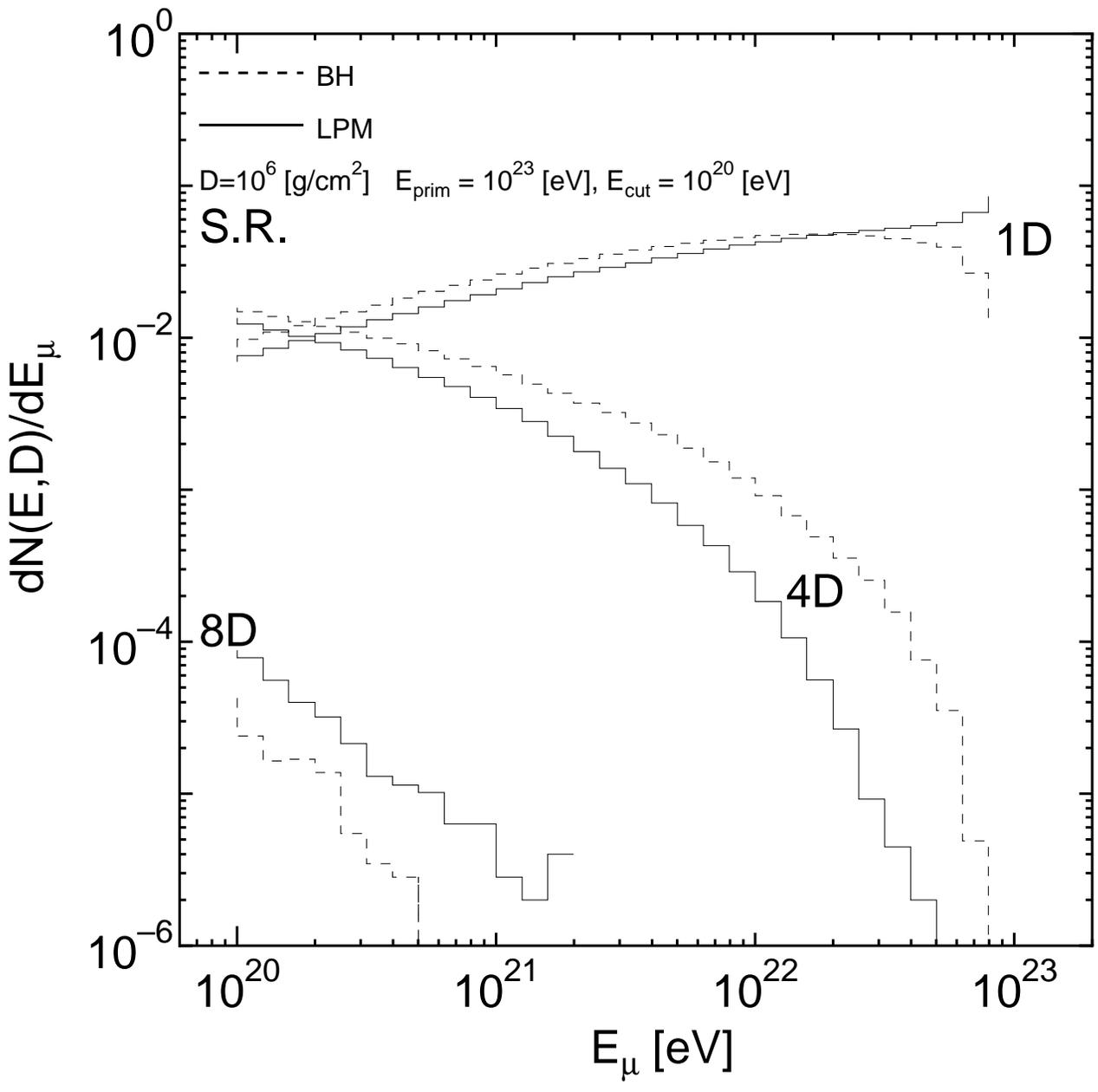
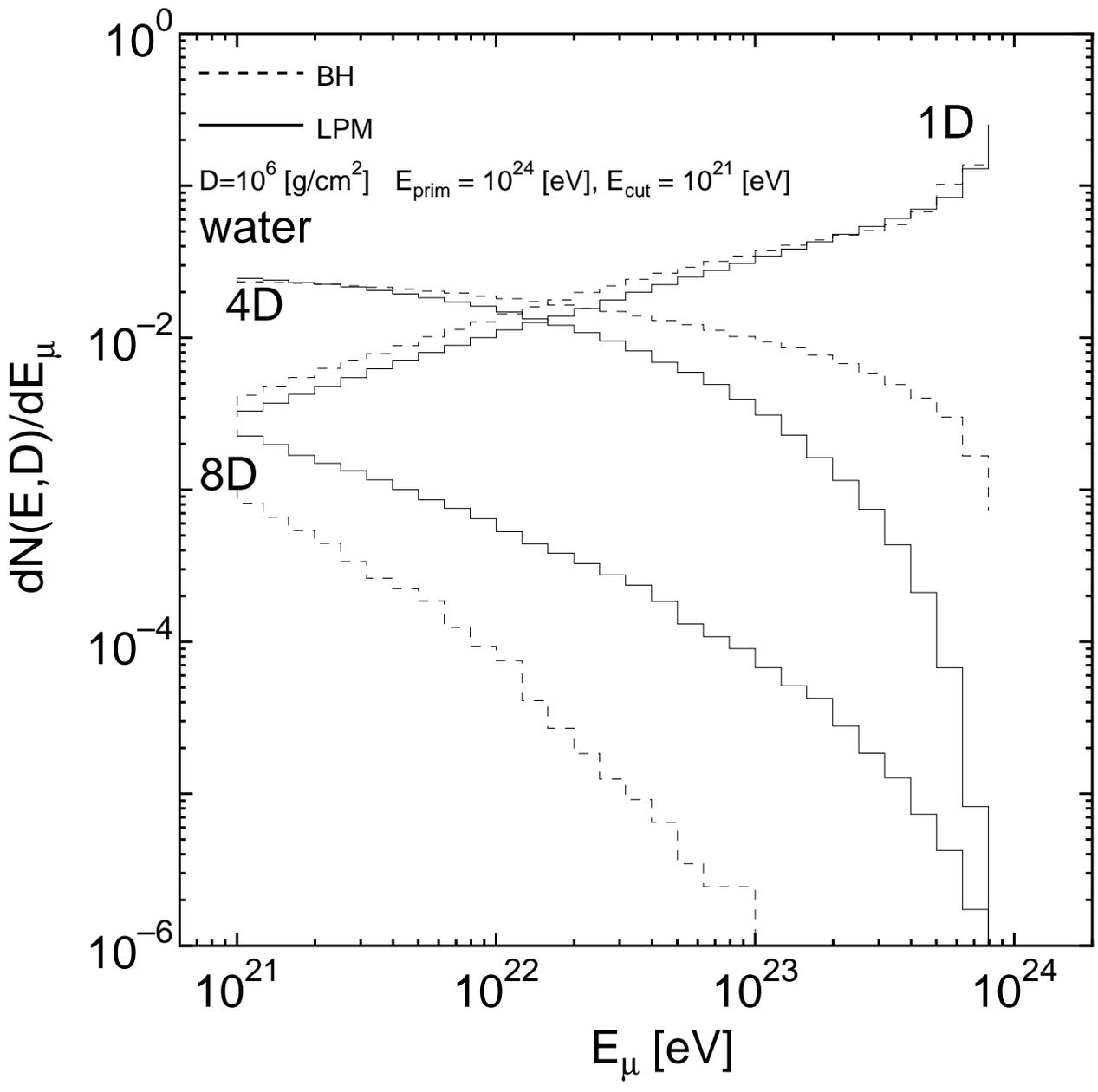
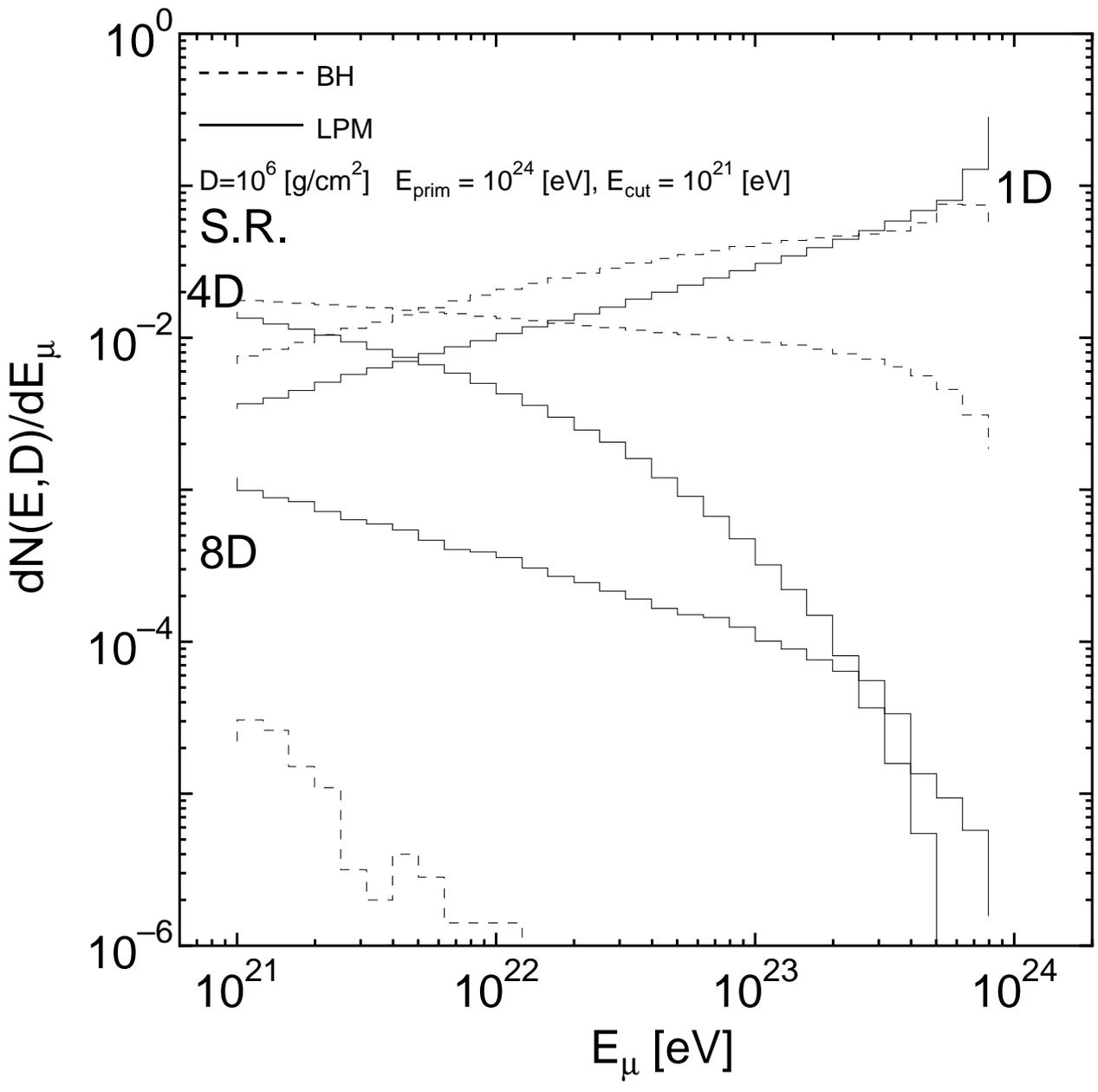
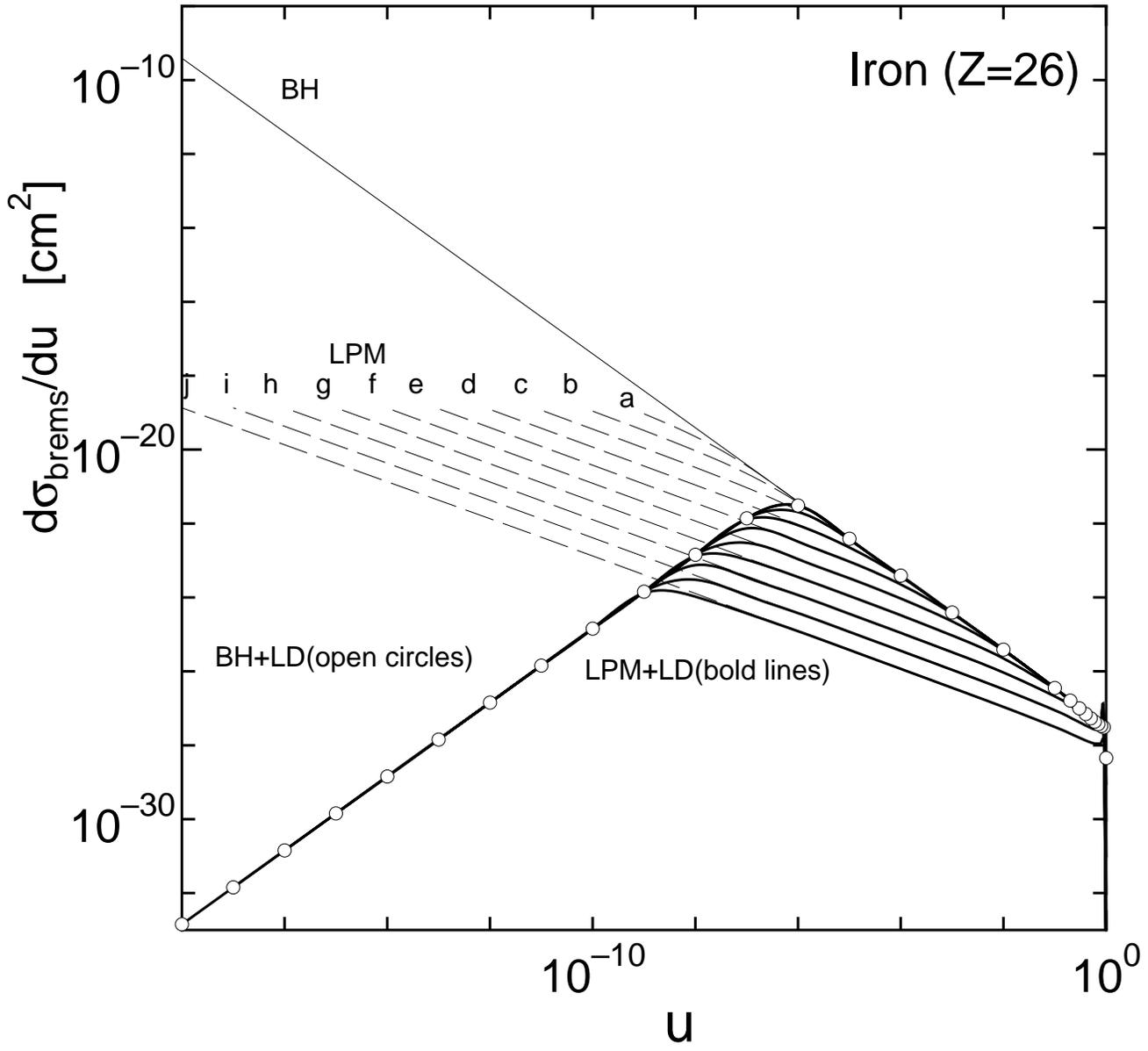
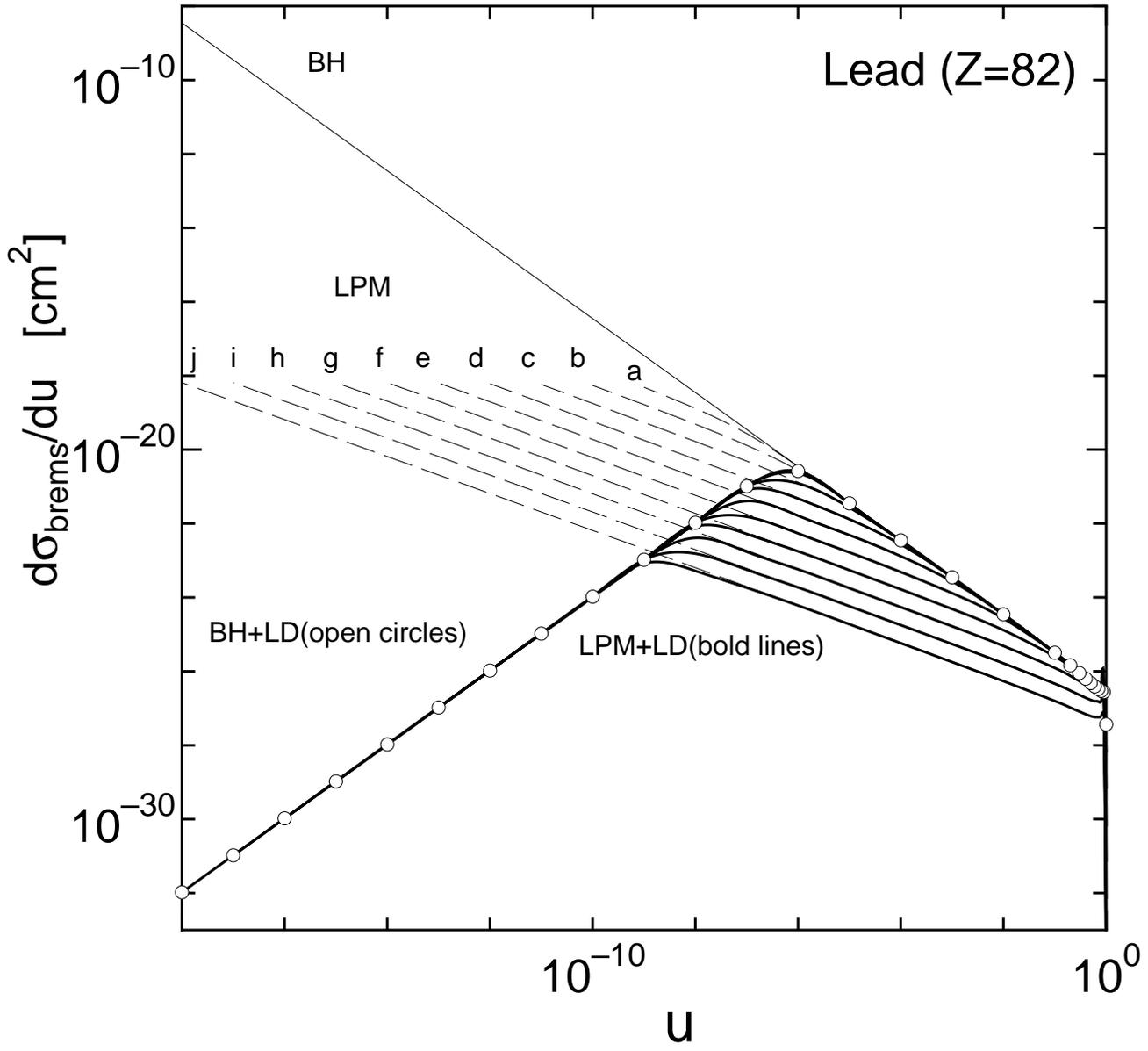
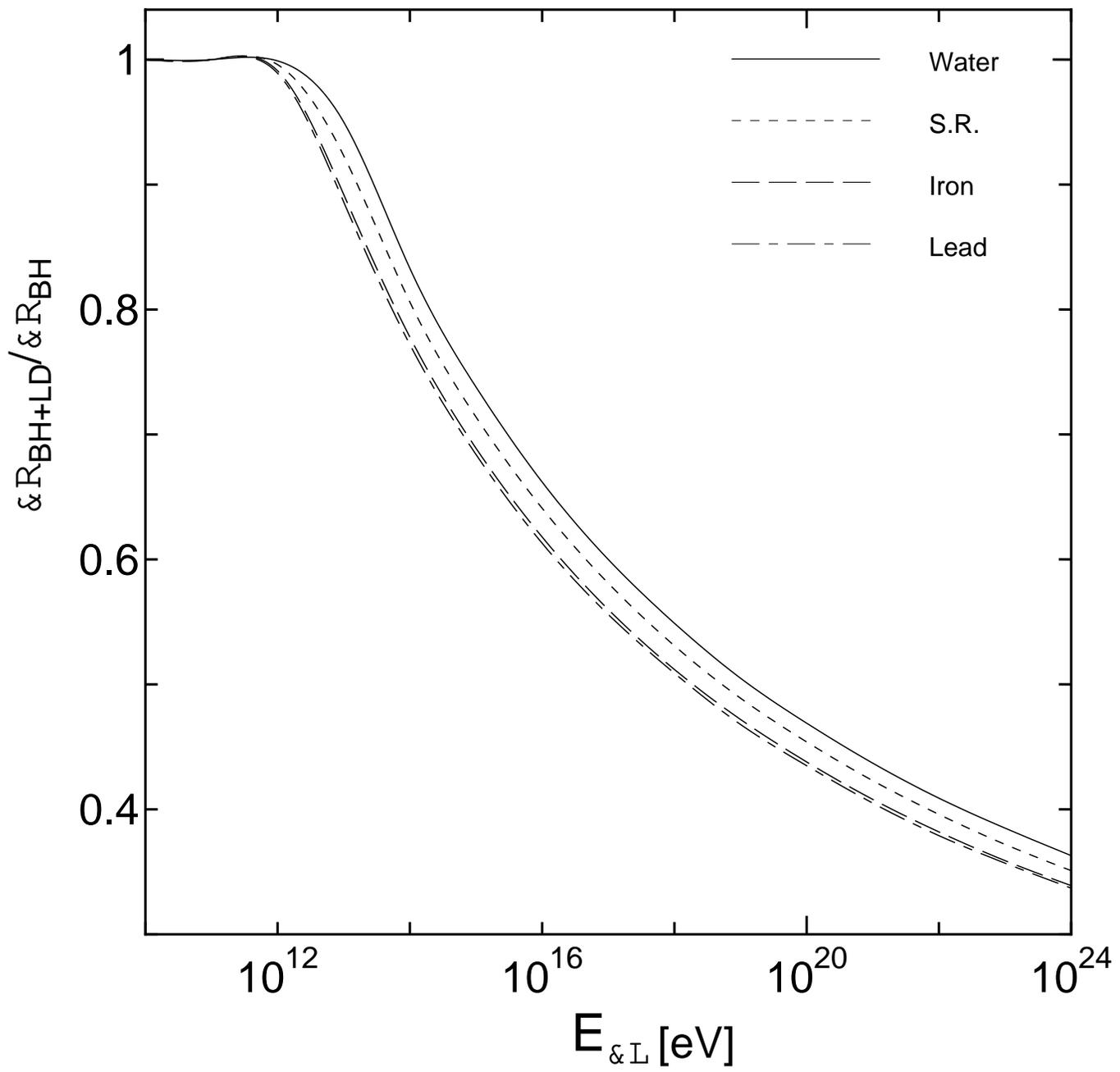
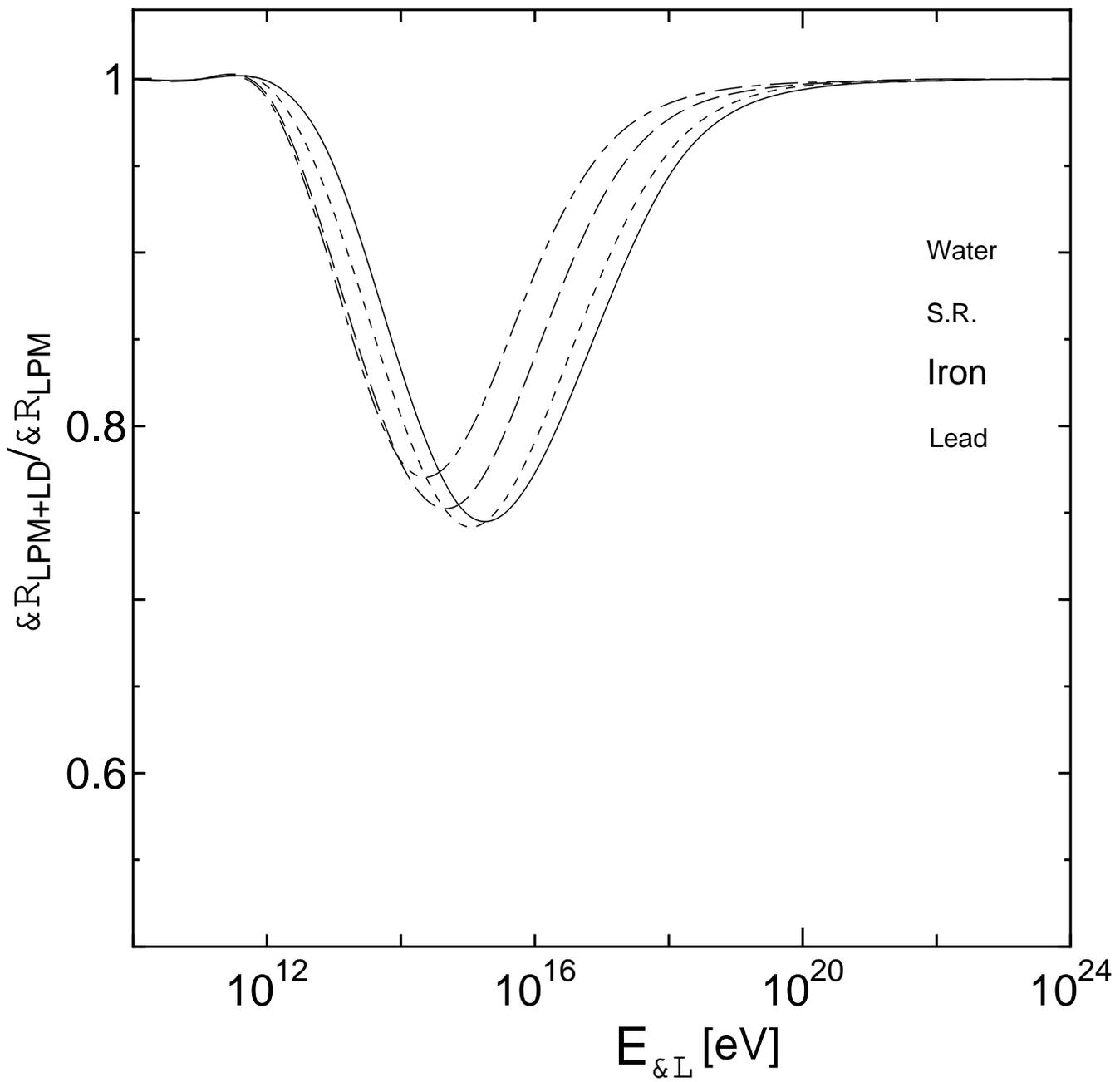
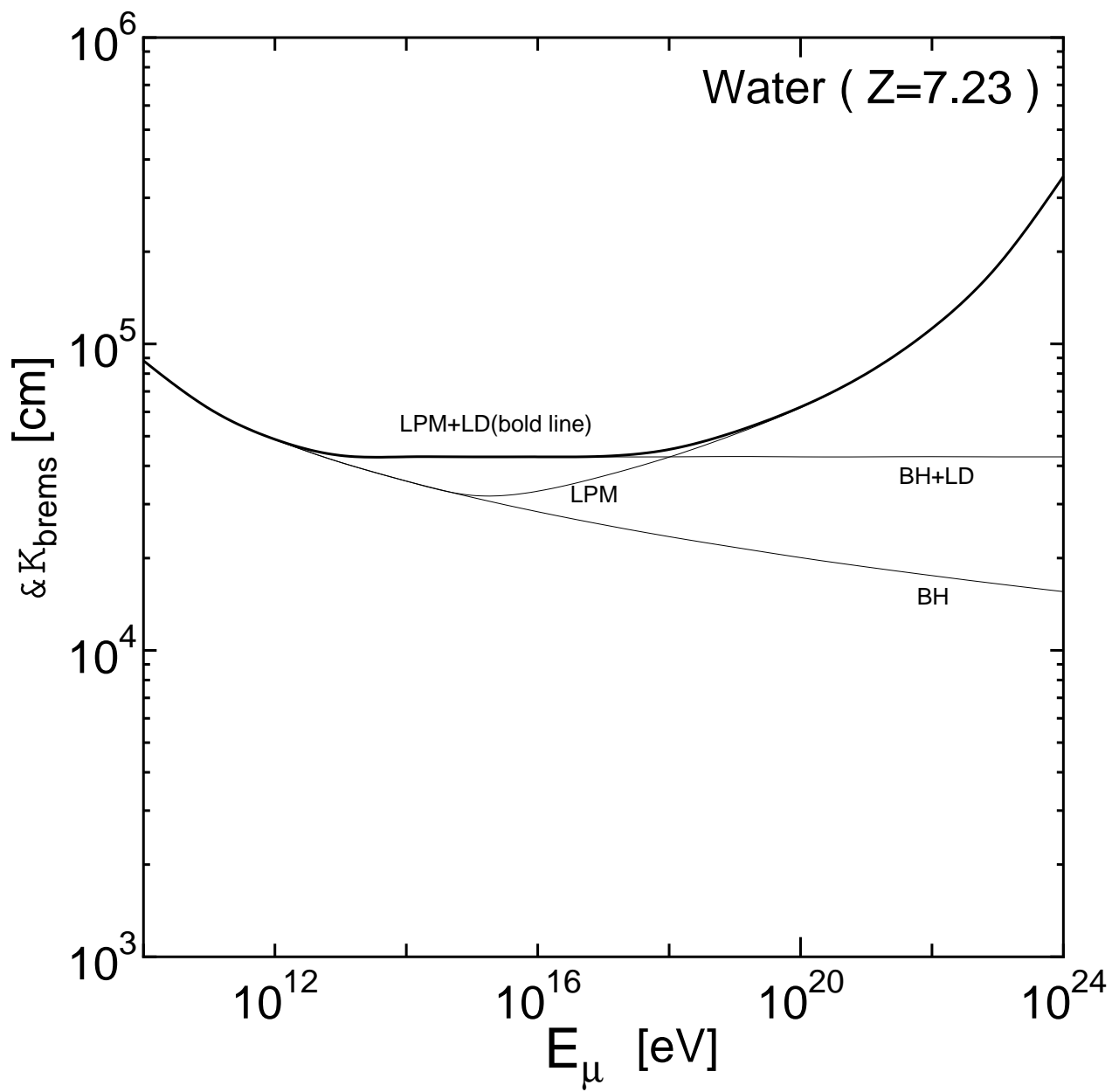
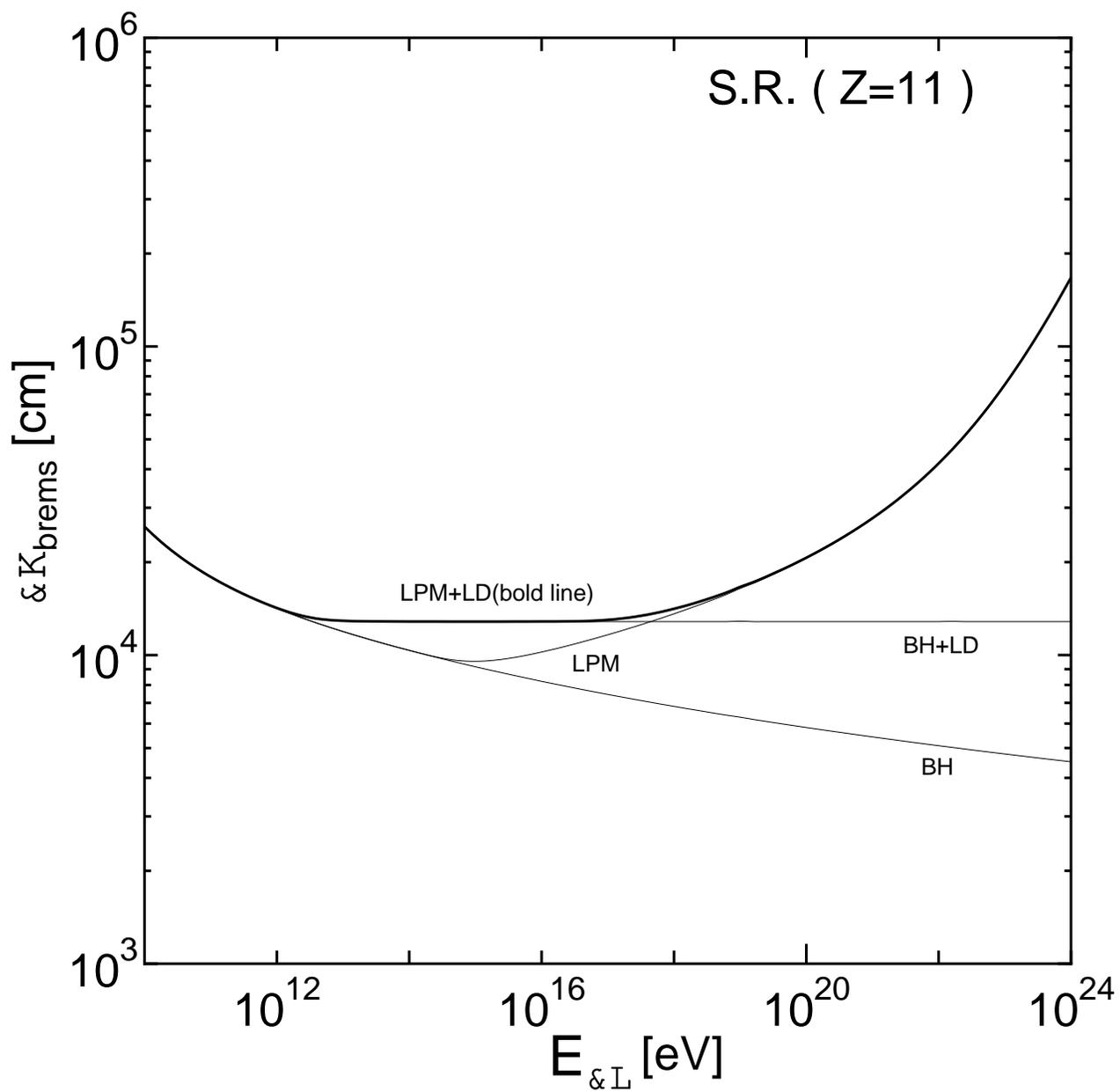
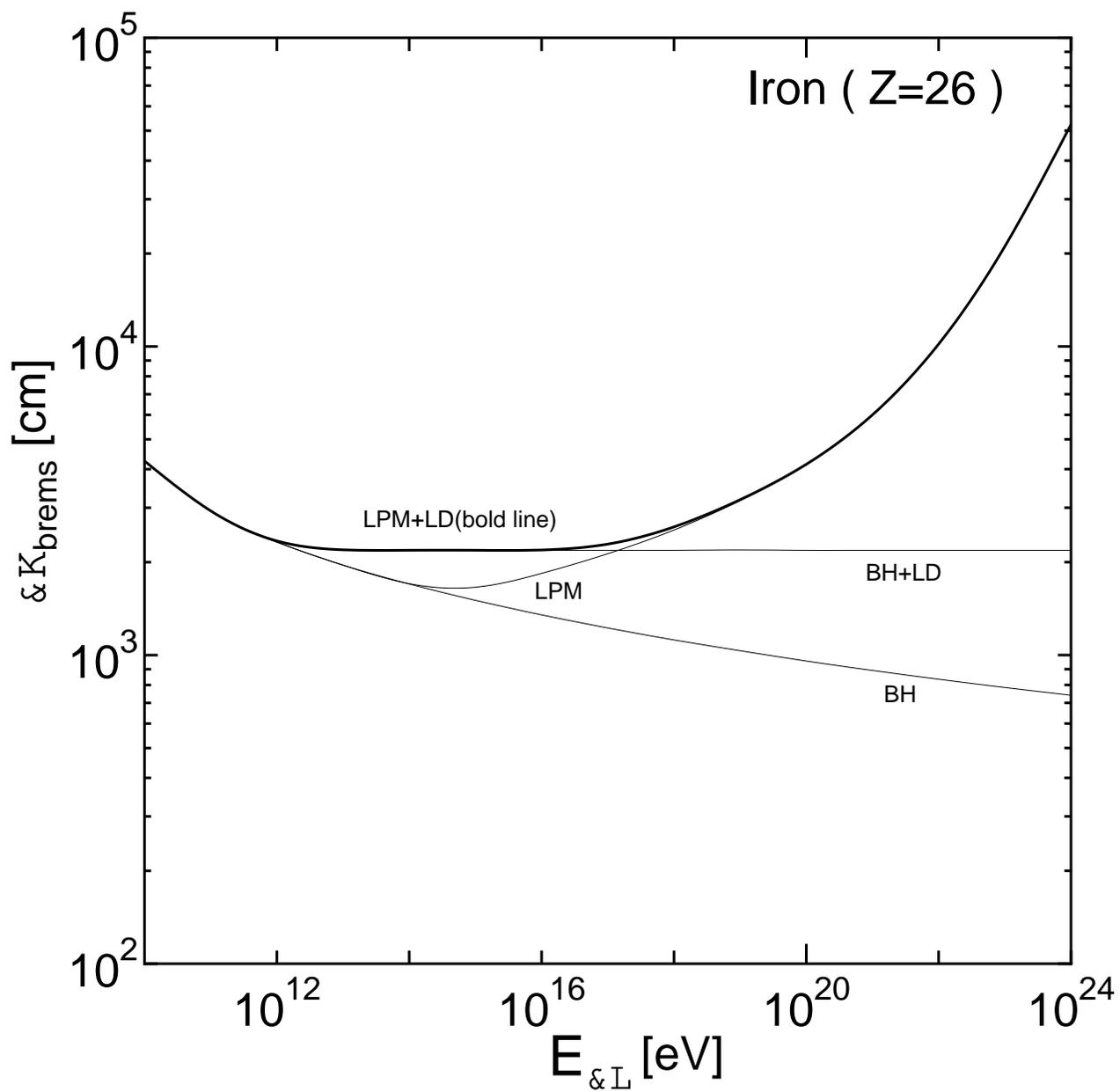
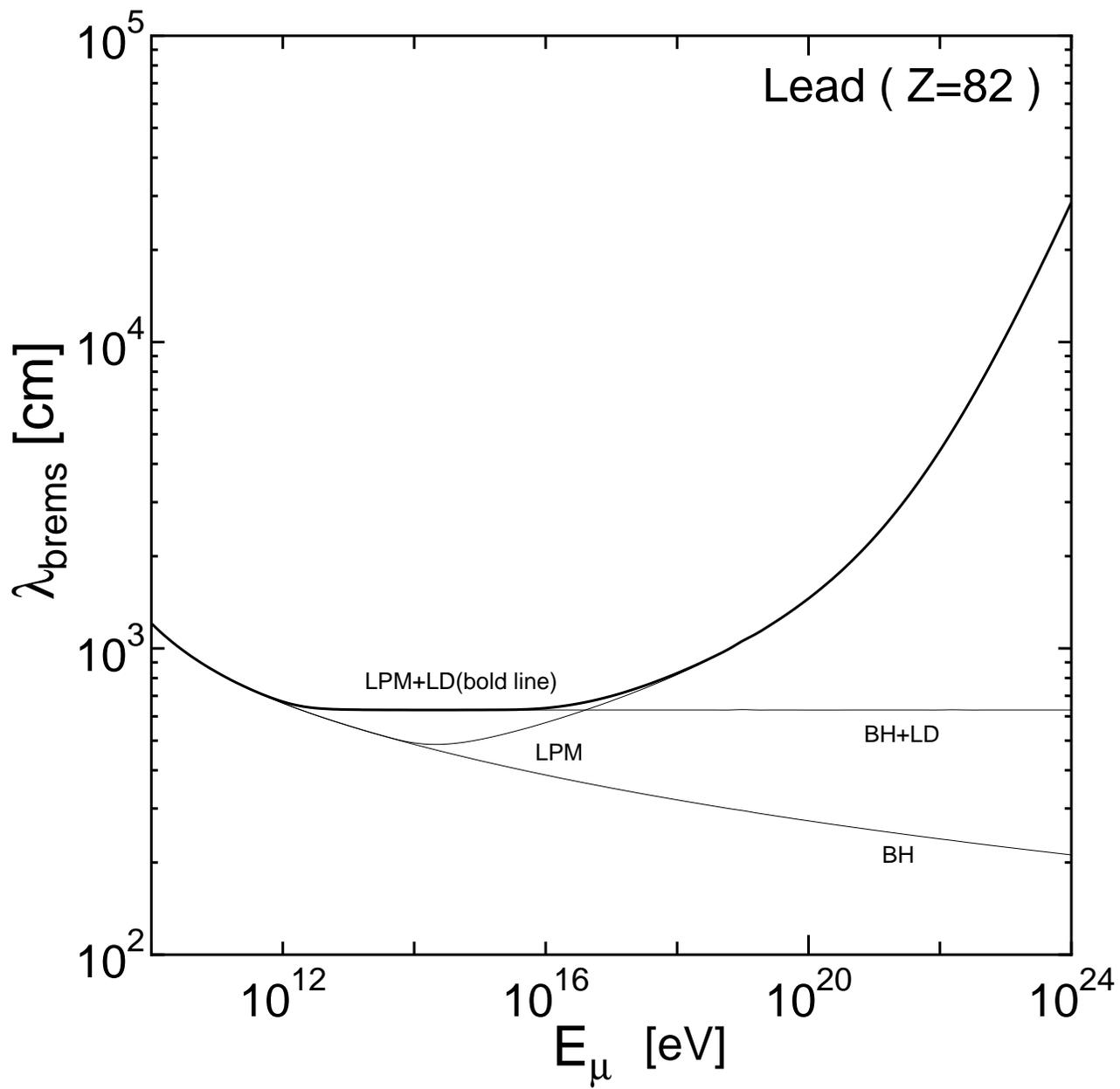
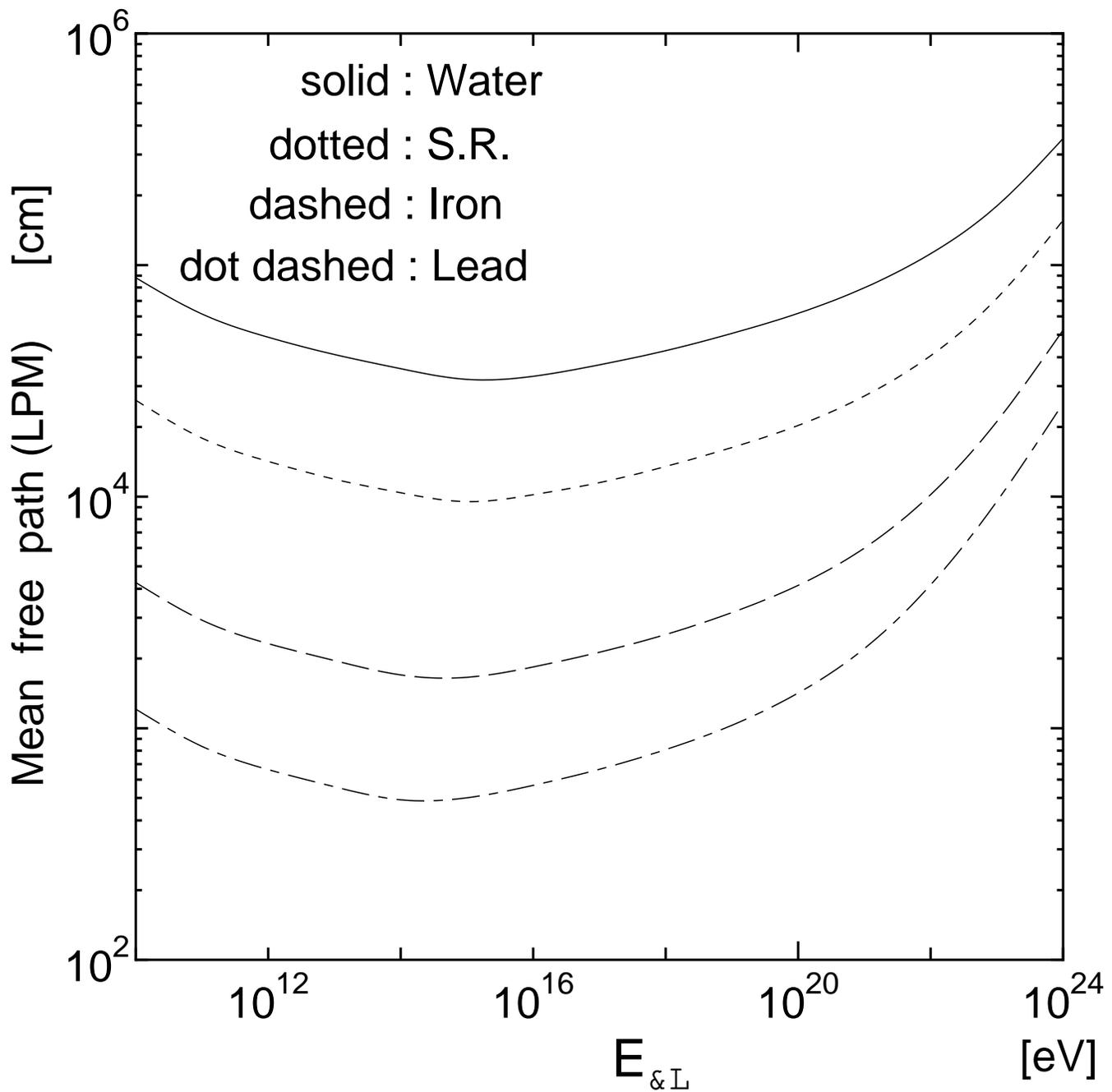
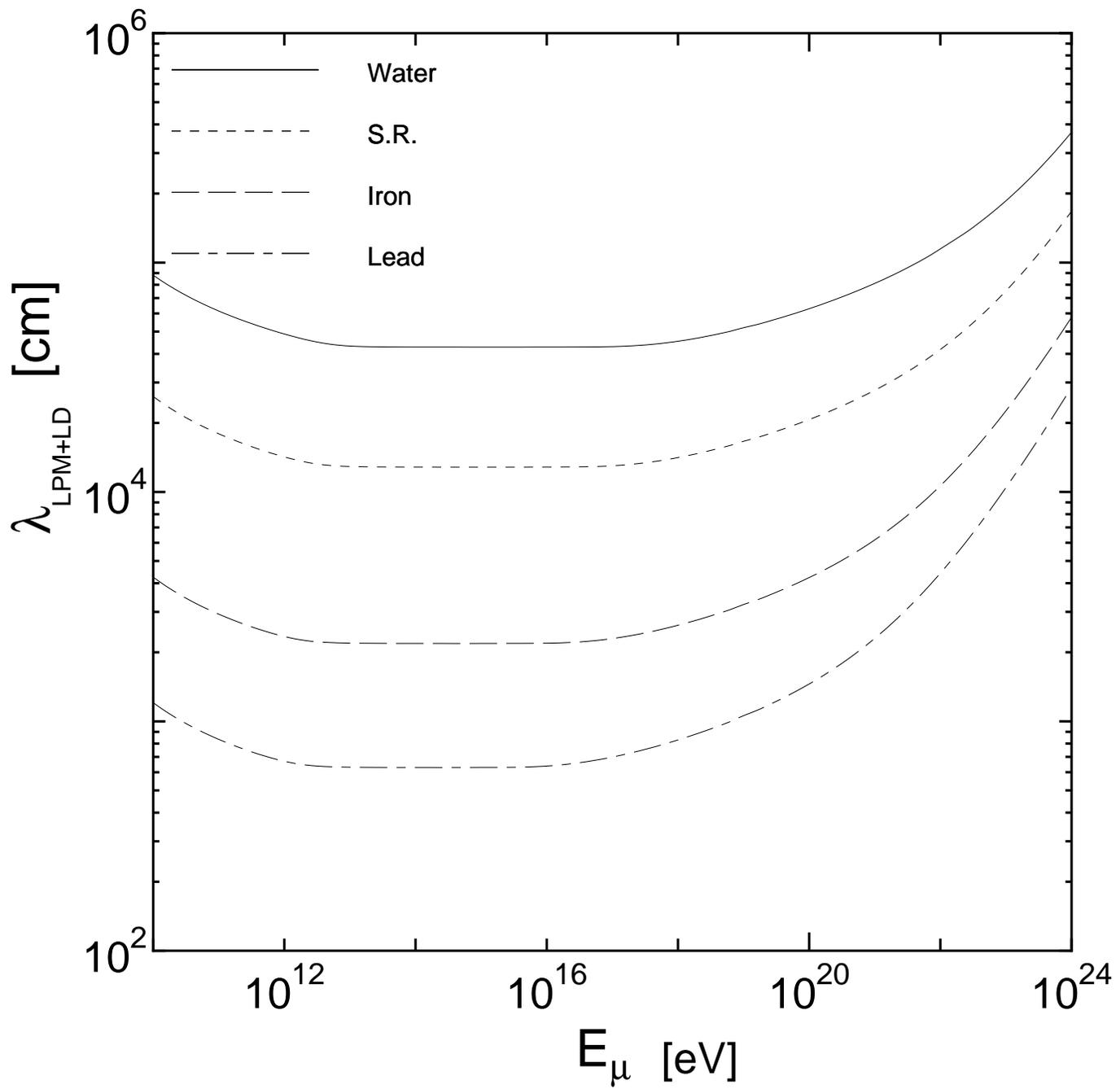
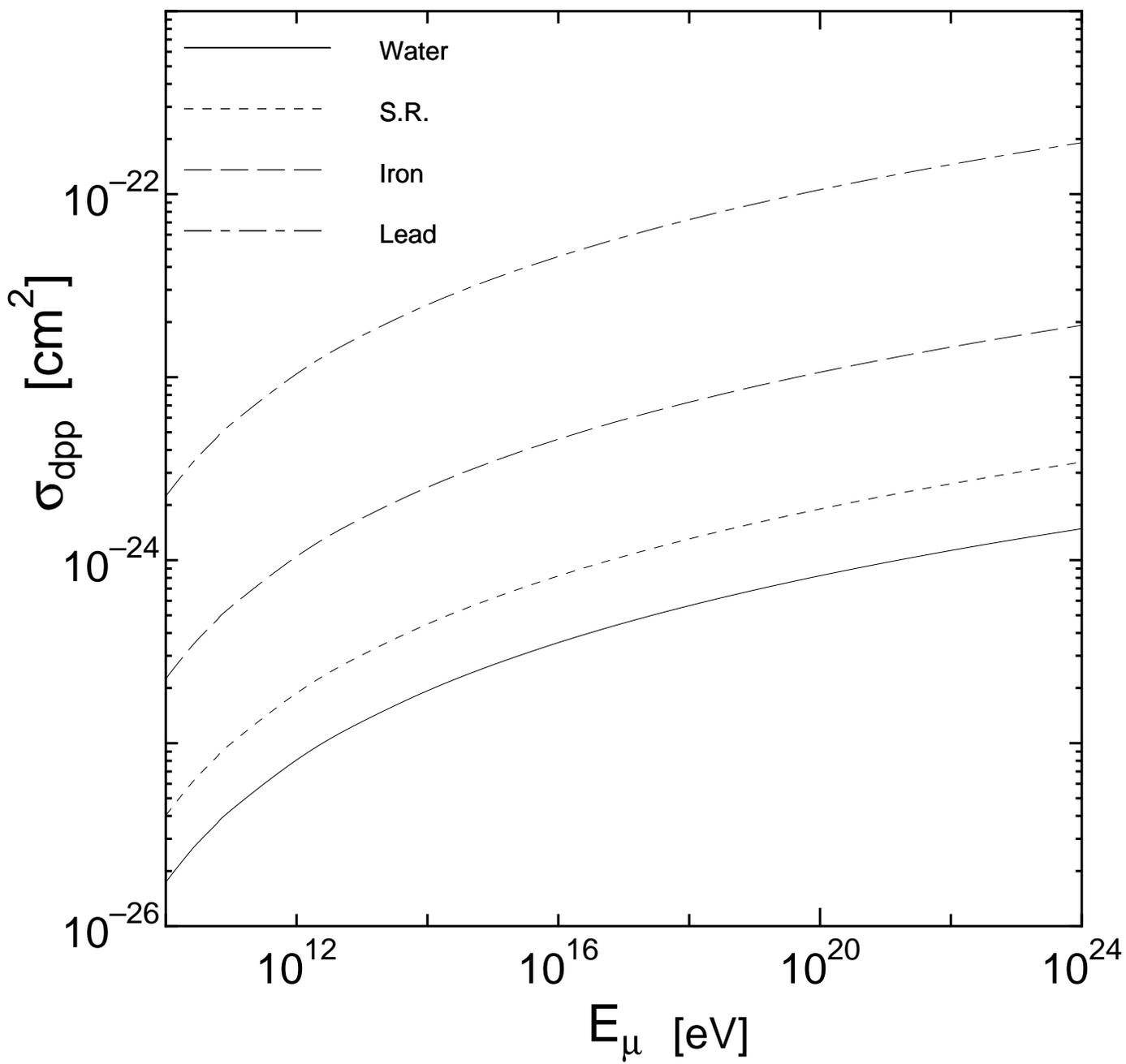
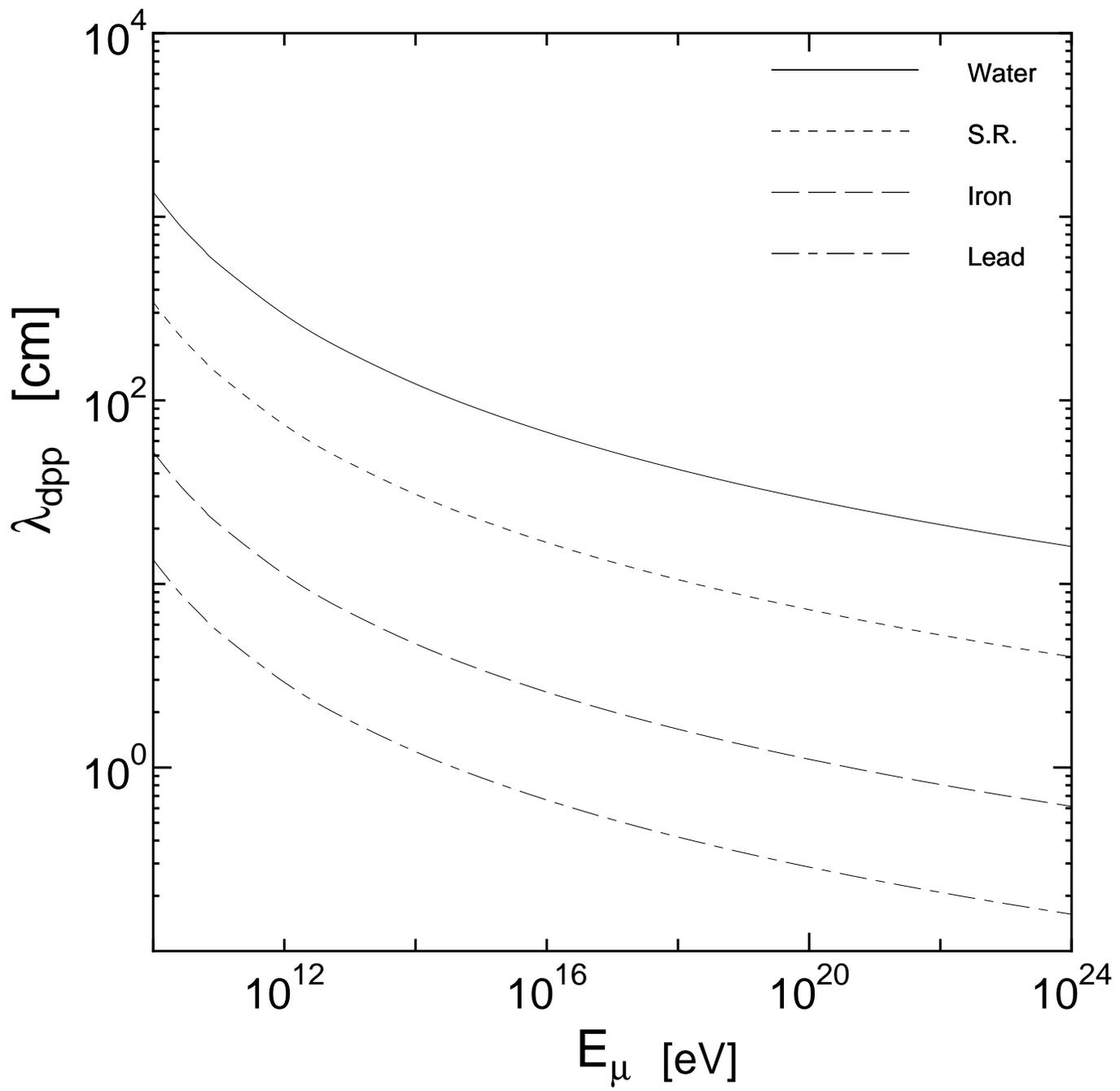
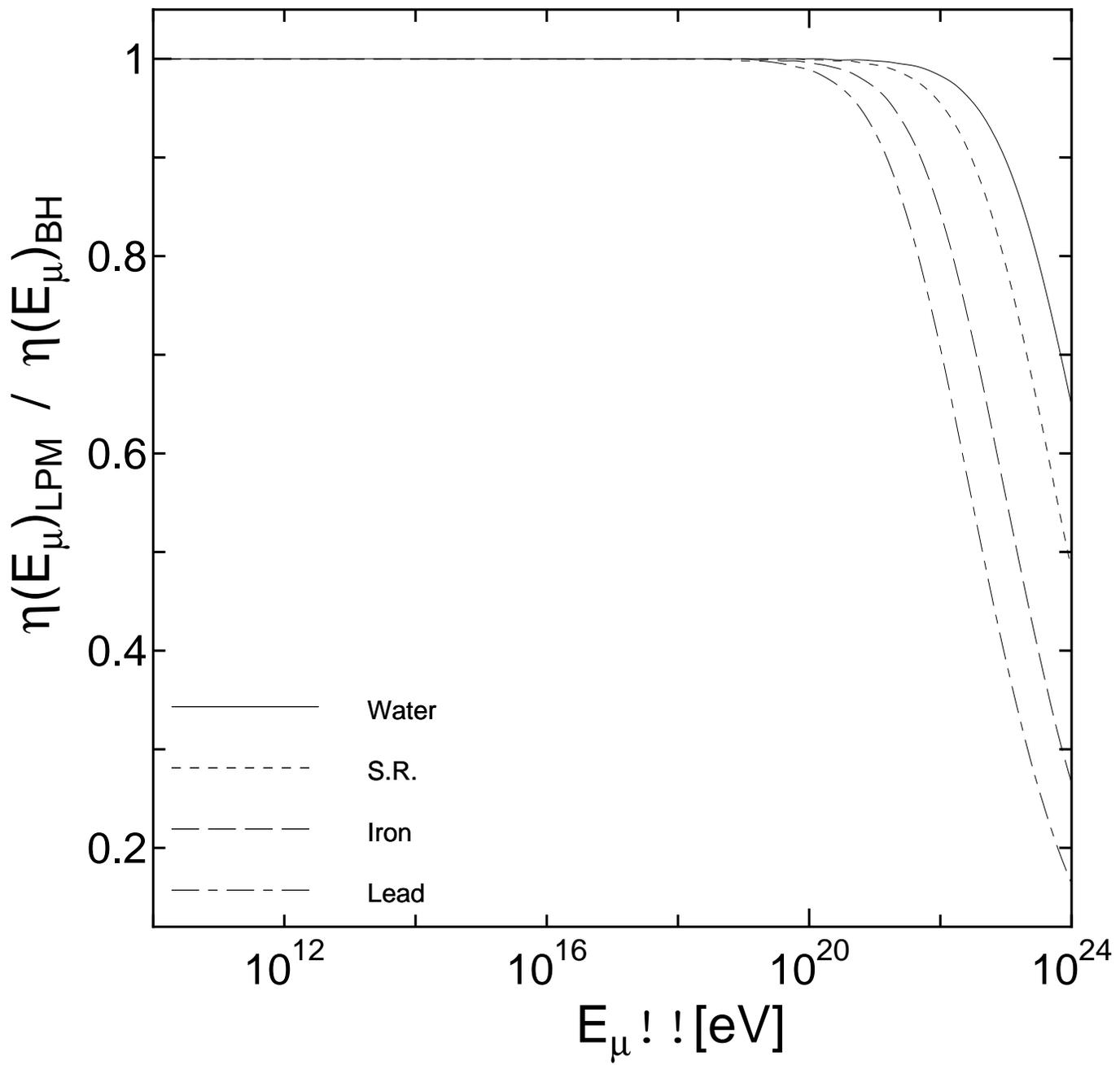
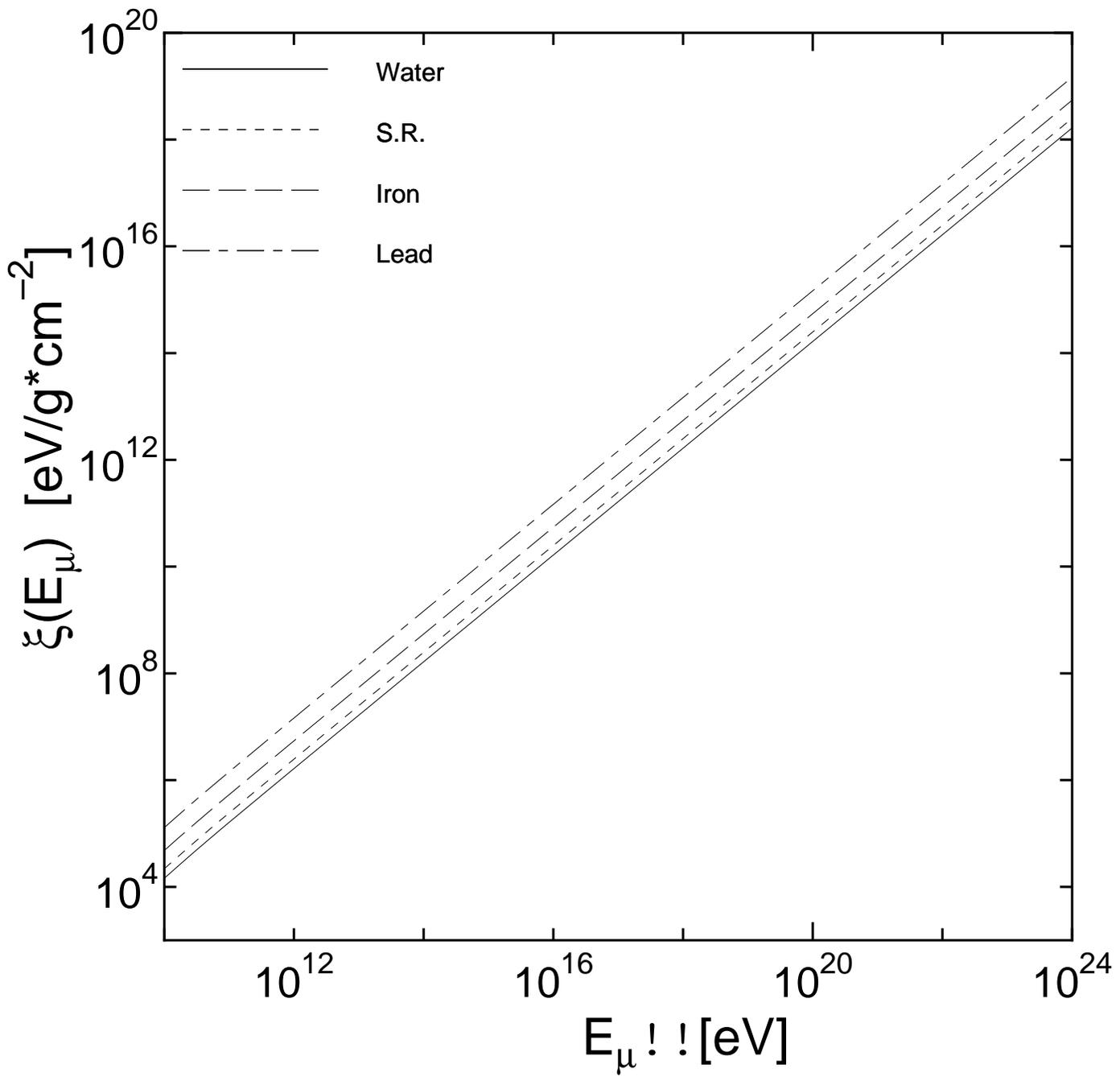
\end{document}